\documentclass[12pt]{iopart}

\expandafter\let\csname equation*\endcsname=\relax 
\expandafter\let\csname endequation*\endcsname=\relax 
\usepackage{iopams,graphicx}  
\usepackage{color,cite,amsmath}
\usepackage[version=4]{mhchem}

\begin{document}

\title[Holey-axicon for free electron laser]{All femtosecond optical pump and X-ray probe: holey-axicon for free electron laser}

\author{V. Anand$^1$, J. Maksimovic$^1$, T. Katkus$^1$, S. H. Ng$^1$, O. Ul\v{c}inas$^2$, \\\protect M. Mikutis$^2$, J. Baltrukonis$^{2,3}$, A. Urbas$^2$, G. \v{S}lekys$^2$, \\\protect H. Ogura$^{4}$, D. Sagae$^{4}$,  T. Pikuz$^{5,6}$, T. Somekawa$^7$, N. Ozaki$^{4,8}$, A. Vailionis$^{9,10}$, G. Seniutinas$^{11}$, V. Mizeikis$^{12}$,  K. Glazebrook$^{13}$, J. P. Brodie$^{13}$, P. R. Stoddart$^{14}$,  L. Rapp$^{15}$, A. V. Rode$^{15}$,\\\protect  E. G. Gamaly$^{15}$, S. Juodkazis$^{1,16}$ 
}

\address{$^1$~Optical Sciences Centre and ARC Training Centre in Surface Engineering for Advanced Materials (SEAM), School of Science, Swinburne University of Technology, Hawthorn, VIC 3122, Australia\\
$^2$~Altechna R\&D, Mokslinink\c{u} st. 6A, 08412 Vilnius, Lithuania\\
$^3$~Laser Research Center, Physics Faculty, Vilnius University, Sauletekio Ave. 10, Vilnius, Lithuania\\
$^4$~Graduate School of Engineering, Osaka University, Suita, Osaka 565-0871, Japan Yamada-oka, Suita, Osaka 565-0871, Japan \\
$^5$~Institute for Open and Transdisciplinary Initiatives, Osaka University, Suita, Osaka, 565-0871, Japan\\
$^6$~Joint Institute for High Temperatures, RAS, Moscow, 125412, Russia\\
$^7$~Institute for Laser Technology, Nishi-ku, Osaka 550-0004, Japan\\
$^8$~Institute of Laser Engineering, Osaka University, Suita, Osaka 565-0871, Japan Yamada-oka, Suita, Osaka 565-0871, Japan\\
$^9$~Stanford Nano Shared Facilities, Stanford University, Stanford, CA 94305, USA\\
$^{10}$~Department of Physics, Kaunas University of Technology, LT-51368 Kaunas, Lithuania\\
$^{11}$~Paul Scherrer Institute, Villigen CH-5232, Switzerland\\
$^{12}$~Research Institute of Electronics, Department of Electronics and Materials Science, Graduate School of Engineering, Shizuoka University, 3-5-1 Johoku Naka-ku
Hamamatsu 432-8011, Japan\\
$^{13}$~Centre for Astrophysics and Supercomputing, School of Science, Swinburne University of Technology, Hawthorn, VIC 3122, Australia\\
$^{14}$Faculty of Science, Engineering and Technology, Swinburne University of Technology, John Street, Hawthorn, Victoria 3122, Australia\\
$^{15}$~Laser Physics Centre, Research School of Physics, The Australian National University, Canberra, ACT 2601, Australia\\
$^{16}$~Tokyo Tech World Research Hub Initiative (WRHI), School of Materials and Chemical Technology, Tokyo Institute of Technology, 2-12-1, Ookayama, Meguro-ku, Tokyo 152-8550, Japan}
\ead{For correspondence: andrei.rode@anu.edu.au  (A.V.R.); sjuodkazis@swin.edu.au (S.J.) }
\vspace{10pt}
\begin{indented}
\item[17]~ May 2020
\end{indented}
\newpage
\begin{abstract}
We put forward a co-axial pump(optical)-probe(X-rays) experimental concept and show performance of the optical component. A Bessel beam generator with a central 100~$\mu$m-diameter hole (on the optical axis) was fabricated using femtosecond (fs) laser structuring inside a silica plate. This flat-axicon optical element produces a needle-like axial intensity distribution which can be used for the optical pump pulse. The fs-X-ray free electron laser (X-FEL) beam of sub-1~$\mu$m diameter can be introduced through the central hole along the optical axis onto a target as a probe. Different realisations of optical pump are discussed. Such optical elements facilitate alignment of ultra-short fs-pulses in space and time and can be used in light-matter interaction experiments at extreme energy densities on the surface and in the volume of targets. Full advantage of ultra-short 10~fs X-FEL probe pulses with fs-pump(optical) opens an unexplored temporal dimension of phase transitions and the fastest laser-induced rates of material heating and quenching. A wider field of applications of fs-laser-enabled structuring of materials and design of specific optical elements for astrophotonics is presented.      
\end{abstract}

%
\vspace{2pc}
\submitto{JPhys Photonics}
%
%
%

\tableofcontents\newpage
\section{Introduction}

This is a concept and perspective paper that discusses the readiness of two different technologies for a new era in ultra-fast femtosecond (fs) pump-probe experiments. We highlight the need for new optical elements that can accommodate a simpler handling of fs-pump and fs-probe beams at vastly different wavelengths: an optical fs-pump at visible $\sim 530$~nm  and X-ray fs-probe at 0.12398~nm (10~keV)~\cite{David}.
The temporal and spatial overlap of ultra-short pulses is always a challenge, especially when the focal spot is $< 10~\mu$m, while the axial extent of a $~\sim 50$~fs pulse is only $15~\mu$m. This is an even greater challenge when the beams are propagating at an angle $\theta$ to each other towards the interaction region, which makes the effective interaction zone extend only a fraction (projection) of the length of the pulse $l_i=d\cos(\theta/2)$ (along the axis of symmetry) depending on their diameter $d$. A front tilt of the beam can be introduced to have the maximum overlap as it would be for the $\theta = 0$ case~\cite{01apl725} for the interfering fs-pulses at large angles using diffractive optical elements (DOEs). 

Nonlinear optical interactions are usually used to determine the temporal overlap of the optical pulses. At high intensity and irradiance (the time average flow of energy per unit time per unit area [W/cm$^2$]) of optical or X-ray pulses, when structural modification of the focal region occurs by ablation, phase transition, and/or color center formation in dielectrics, high precision spatial and temporal overlap of pulses is simplified and achieved by optical monitoring of the interaction zone. Transmitted and reflected beams are used for the final optimisation. In the case of hard X-rays ($\sim 10$~keV) which have a small absorption cross section within materials, timing with fs-laser pulses (optical) can be achieved by detection of increased transmissivity of GaAs under fs-X-ray pulse~\cite{Sato15}. A LiF crystal was used for \emph{in situ} control of  focal position of few-keV  X-ray free electron laser (X-FEL)~\cite{fel} beam and for measurement of a transverse beam intensity profile with a spatial resolution of 1~$\mu$m~\cite{Pikuz}. 

There is an inherent benefit to use axially extended beams such as Bessel-Gauss beams rather those with a Gaussian cross-section. For a co-linear propagation of optical and X-ray pulses their interaction volume can be augmented and spatial-temporal alignment simplified. Obviously, a central hole in an optical element is required for on-axis propagation of the fs-X-ray pulse. Since the focus of a fs-X-ray free electron laser (X-FEL) is tunable by the placement of the sample along the beam with the smallest spot area of $1~\mu$m$^2$ (at Spring-8 facility; Spring-8 Angstrom Compact free electron Laser, referred to as SACLA), a hole of 100~$\mu$m diameter would meet the requirements of alignment tolerances. The optical fs-pump will be delivered onto the target by a flat axicon lens $\sim 4$-mm-diameter, which is typical for free-space beam of most of fs-lasers, and is inscribed around the central hole. This concept of an axially elongated Bessel beam intensity delivered by the on-axis converging conical wave can have several realisations using refocusing and demagnification with a 4f-optical system (see discussion further in the text) or reflective optics to facilitate actual optical-pump X-ray-probe experiments. Recent developments in fs-laser fabrication of flat optical elements have produced a mature technology which is utilised for this demonstration of a new optical element inscribed in a typical pure-silica blank of 3~mm thickness.   

\begin{figure}[tb]
\centering \includegraphics[width=11cm]{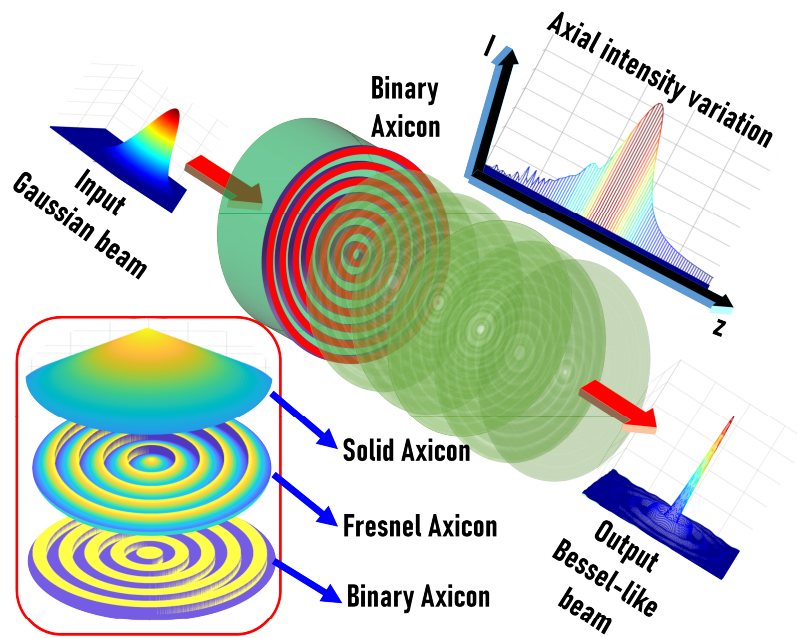}
\caption{A Gaussian beam is converted into a Bessel-like beam using a circular grating, solid axicon, Fresnel or binary axicon; see details in the text.} \label{Axicon}
\end{figure}

Here, we show the principle of needle-beam intensity generation by using flat optics by fs-laser inscription, which is a feasible method for fabrication of optical elements. Then a Bessel-beam generator is demonstrated by controlled patterning with nanoscale form-birefringence. The very same fs-laser inscription was used for fabrication of the central hole in the optical element by fs-laser induced breakdown and chemical wet etching~\cite{01ol277,04apa1549,06apa99}. The fabricated optical element is characterised and discussion of its applicability for X-FEL experiments is presented, based on a recent demonstration of ns-pump (optical) and fs-probe (X-FEL) study of laser-induced compression of corundum~\cite{Inubushi}. We discuss the prospects for probing matter under "warm-dense" conditions~\cite{wdm1,wdm2} when the potential energy of the interaction between electrons and nuclei and the kinetic energy of electrons are of the same magnitude (as in the interior of planets). Ultra-fast solid-liquid-gas-plasma transitions and subsequent quenching of materials into new exotic phases can be monitored with unprecedented temporal resolution using X-ray diffraction (XRD)~\cite{Inubushi}.    

\section{Making optics flat for faster laser inscription}

One obvious simplification of lens fabrication and increase in the fabrication throughput is to make optics flat by inscription of the required refractive index pattern for the light-ray bending function (e.g., a curved surface of a lens). Combination of refractive and diffractive functions can be used for specific optical elements, e.g., the function of a conical lens (axicon) can be achieved by a Fresnel axicon~\cite{Golub}  or a binary circular grating. The circular grating can convert a Gaussian beam or a uniform illumination into a Bessel-like beam~\cite{Vijayakumar}. The conversion of a Gaussian beam into a Bessel-like beam using a circular grating is shown in Fig.~\ref{Axicon}. A binary axicon is the simplest possible realization of an axicon. A solid axicon and a Fresnel axicon can function either as a refractive element or a diffractive element depending upon the features involved. Contrarily, a binary axicon is a pure diffractive element. This discussion of refractive vs diffractive optics is important in regards to optical losses and efficiency of the optical element when high irradiance has to be reached at the focus (on the target); see Sec.~\ref{con}. 

\begin{figure}[tb]
\centering \includegraphics[width=1\linewidth]{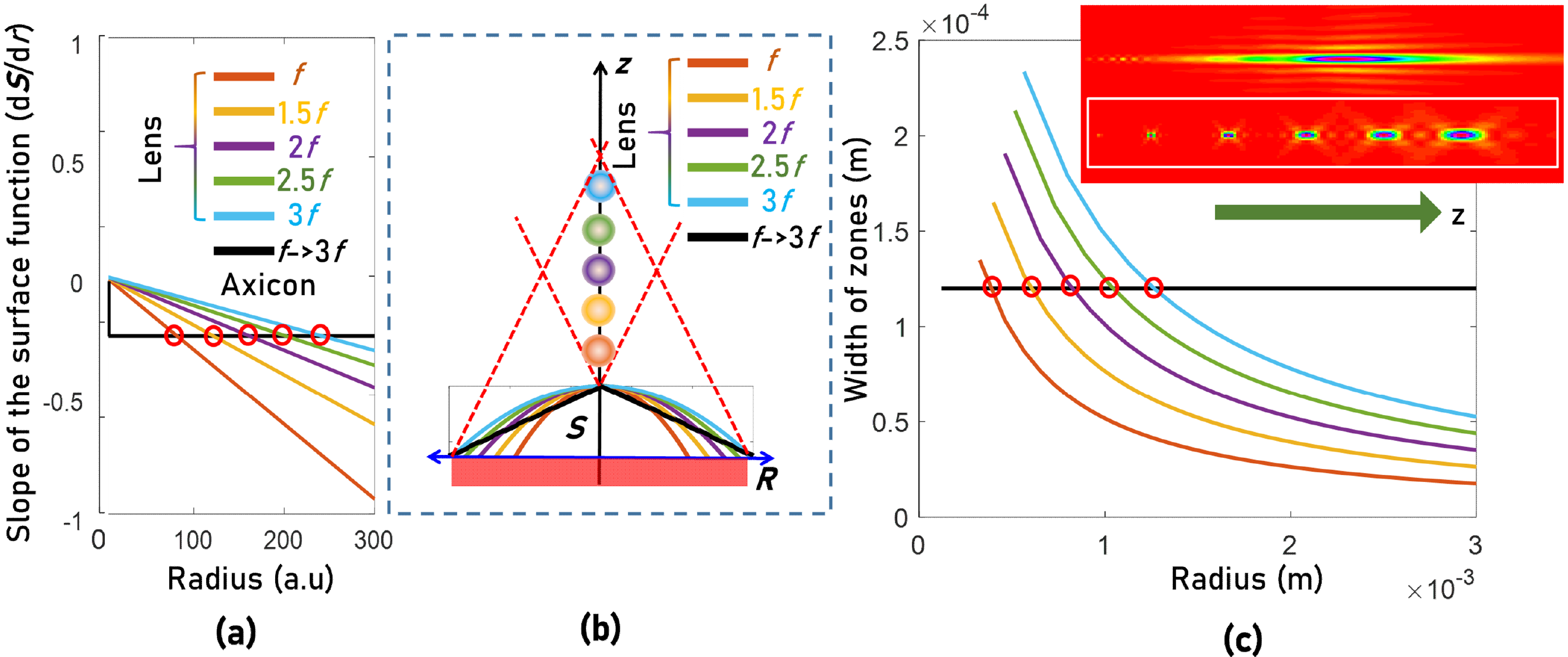}
\caption{(a) Plot showing the derivative of the surface functions of an axicon and lenses with different focal lengths as a function of the radial coordinate. (b) Plot of the surface function $S$ as a function of the radial coordinate. (c) Plot of the width of the zones of Fresnel zone plate and an axicon. Inset shows axial intensity for the axicon and that of the five different lenses with focal lengths $f$ to $3f$ in steps of $0.5f$.} \label{Figure21}
\end{figure}

There are numerous perspectives to understanding light modulation by an axicon~\cite{McLeod,Scott,Kizuka} and a different perspective is presented here. A solid axicon's surface profile is linear with respect to the radial coordinate $R$ given by $S_{1}=- R\alpha$, where $\alpha$ is the base angle and $R=\sqrt{x^2+y^2}$. The surface of a lens is given by $S_{2}=-R^2/2 f$ (paraxial approximation), where $f$ is the focal length of the lens. A lens collects all the incident light at different values of the radial coordinate and focuses it on a point with the entire radial coordinate mapped to a single point in space when $z=f$. Contrarily, an axicon maps every value of the radial coordinate to a different point on the optical axis, focusing light rings with different radii at different locations on the optical axis. The result is a large focal depth and self-reconstructing behaviours~\cite{Bouchal}. In a way, an axicon can be considered as a collection of infinitesimal rings of lenses of different focal lengths stacked along the radial coordinate. To prove this, the two surface functions $S_{1}$ and $S_{2}$ are differentiated $dS/dR$ with respect to the radial coordinate and plotted for five values of focal length of lenses $f$ to $3f$ in steps of $0.5f$ in Fig.~\ref{Figure21}. From the Fig.~\ref{Figure21}(a), it is seen that the slope values of the lenses meet the constant slope value of the axicon at different regions. Stronger(faster) lenses meet sooner than the weaker(slower) ones. The surface functions are also plotted for reference in Fig.~\ref{Figure21}(b).

\begin{figure}[tb]
\centering \includegraphics[width=0.9\linewidth]{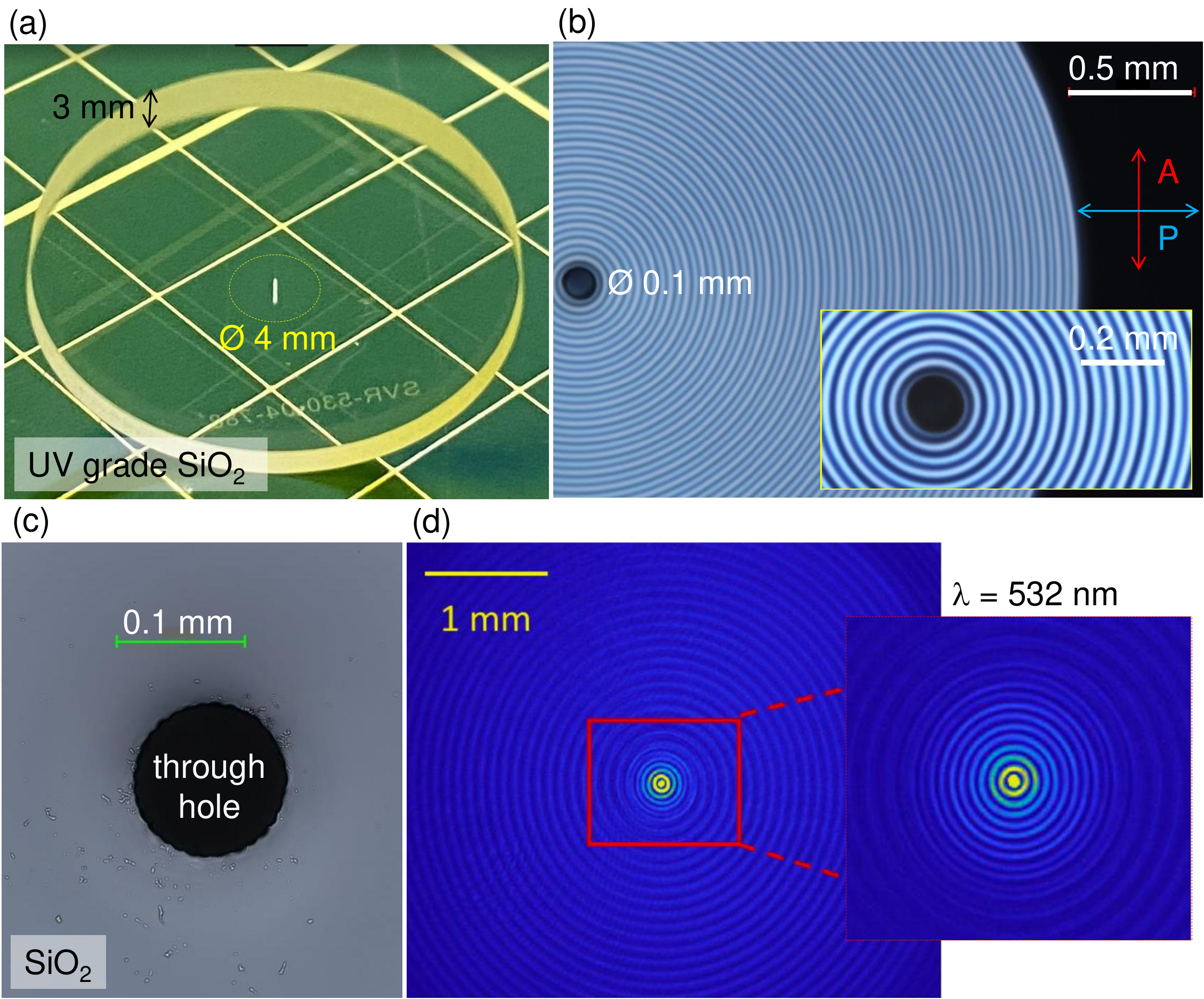}
\caption{(a) Axicon lens inscribed in UV-grade \ce{SiO2} (UVFS) with clear aperture 4~mm, central through hole of $100~\mu$m diameter and corresponds to the refractive axicon with an equivalent apex angle of 178$^\circ$ ($\alpha = 1^\circ$). The retardance in the laser inscribed regions corresponds to $\lambda/2$ for 530~nm. (b) Cross polarised image of the laser inscribed axicon; arrows mark orientations of polariser P and analyser A. (c) Central through-hole chemically etched out after laser inscription. (d) Light intensity distribution at close to maximum of axial intensity. Overall transmission of the lens $T\approx 90\%$ (without anti-reflection coating).  } \label{f-hole}
\end{figure}

\begin{figure}[tb]
\centering \includegraphics[width=0.65\linewidth]{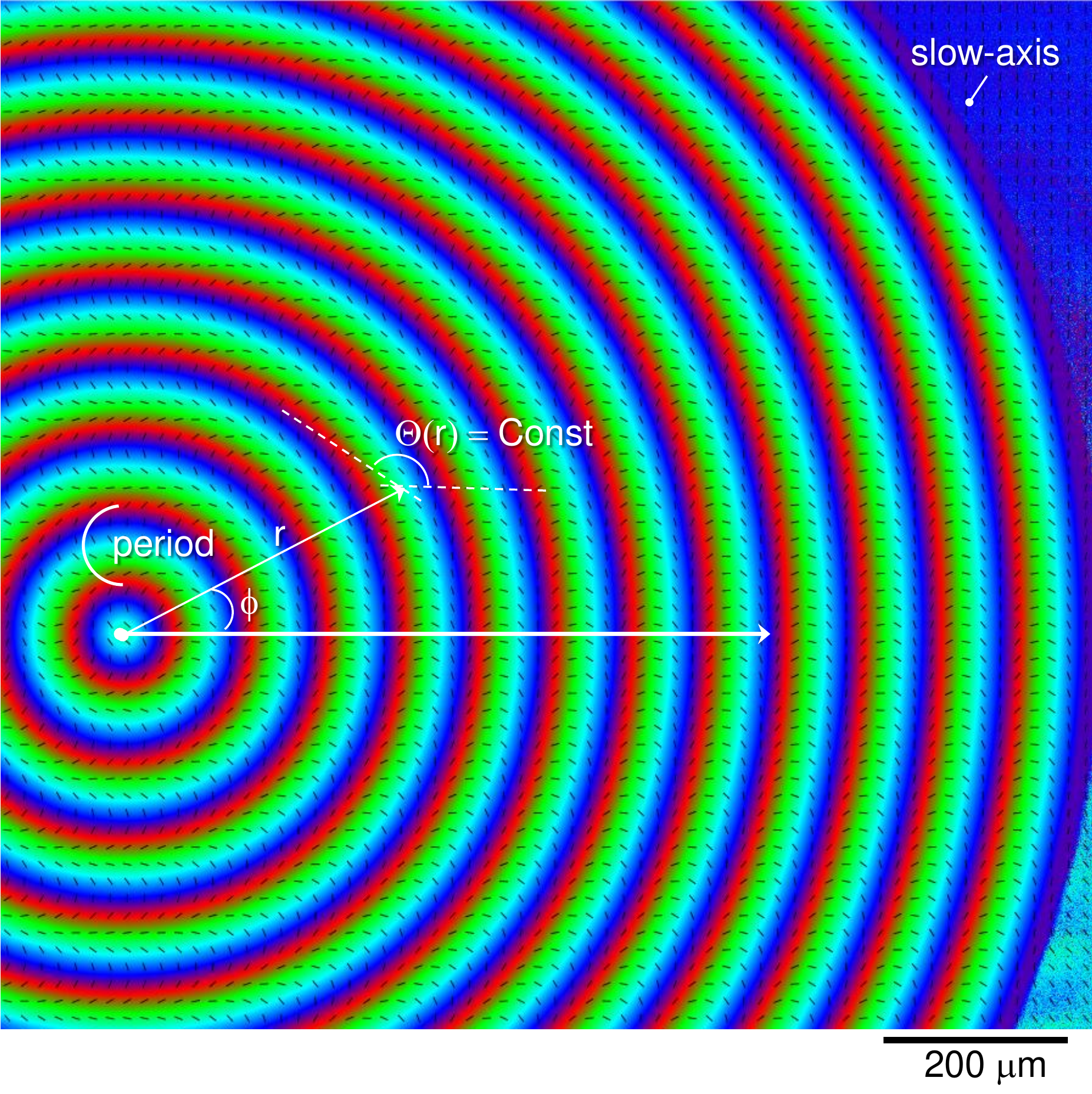}
\caption{Slow-axis orientation map of form-birefringent nanogratings inscribed in silica by fs-laser pulses (blue-to-red covers $0-\pi$ orientation angles of slow-axis). The local-azimuth $\theta$ orientation of slow-axis is repeated periodically in a circular pattern. The orientation of gratings was dependent on radius $r$ as local-azimuth $\theta(r) = Const$ and was not dependent on the polar angle $\phi$.  } \label{f-grat}
\end{figure}

The same analogy applies to diffractive lenses and axicons except that the beam modulation is achieved using diffraction rather than refraction in the above cases~\cite{VijayakumarSPIE}. As the refraction angles are controlled by the surface profile of the refractive elements, the diffraction angles are controlled by the width of the zones. A  diffractive axicon has a fixed period $\Lambda$ while the period of the Fresnel zone lenses vary with the radial coordinate given by $\Lambda_{fzl}=\sqrt{2(m+1)f\lambda}-\sqrt{2mf\lambda}$, where $m$ is order of the zones. Again comparing an axicon's period with that of a Fresnel zone lens (Fig.~\ref{Figure21}(c)), it is seen that the width of the zones meet the period of the axicon at different radial values extending the earlier analysis to diffractive domain. The light modulation by a diffractive axicon and a Fresnel zone lens ($f$ to $3f$ in steps of $0.5f$) were simulated using scalar diffraction formulation with Fresnel approximation as a convolution between the complex amplitude of an axicon given by $exp(-j2 \pi R/\Lambda$) or that of a Fresnel zone lens given by $exp(-j \pi R^2/\lambda f$) with a Quadratic phase function with a complex amplitude $exp(j \pi R^2/\lambda f$) as shown in an inset in Fig.~\ref{Figure21}. When the complex amplitude of a  Fresnel zone lens is convolved with that of a quadratic phase function, both with the same focal distances, a delta-like function is generated indicating that light is focused on a point. On the other hand, when the complex amplitude of an axicon is convolved with the quadratic phase function, a part of the above two functions are conjugates of one another (crossing points in Fig.~\ref{Figure21}(b)) generating a delta light function (central peak of the Bessel function), while the other parts contribute to the rings around the central peak upon convolution. For different axial distances, different radial regions of the axicon matches with the conjugate of the corresponding quadratic phase functions resulting in a central peak with the uncompensated radial regions contributing to the rings formation around the central peak. When the two diffractive functions are binary, then there are slight deviations corresponding to the higher diffraction orders. 

Engineering of refractive index pattern using fs-laser inscription for a specific optical function (e.g., defined by a 3D surface of a lens/axicon) can be used for fabrication of a simple flat lens which is described next.  

\begin{figure}[tb]
\centering \includegraphics[width=0.95\linewidth]{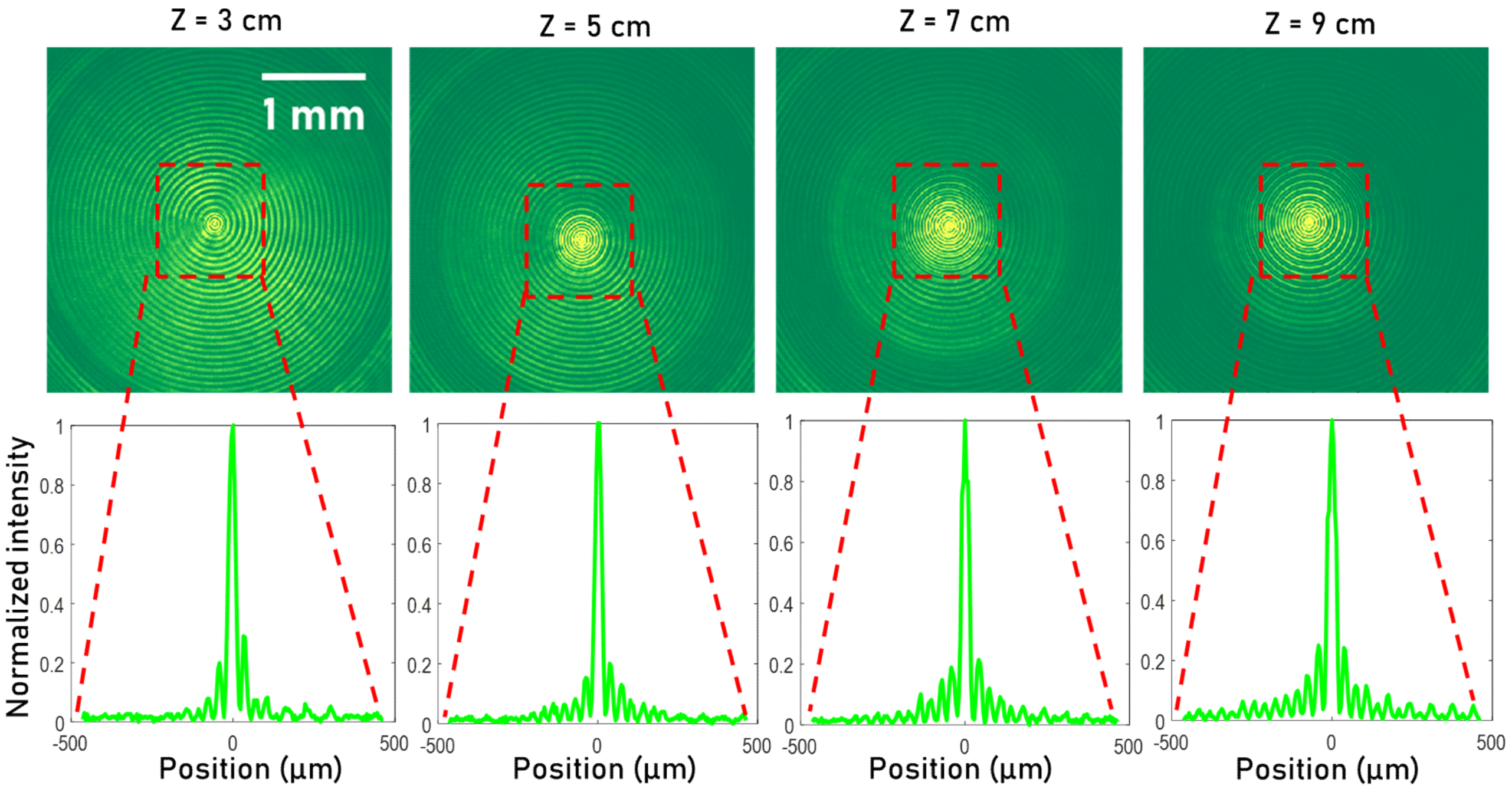}
\caption{Diffraction patterns recorded at $Z = 3$~cm, 5 cm, 7 cm and 9 cm from the axicon. The central spot has diameter of $\sim 37~\mu$m at $1/e^2$-level of intensity;  polarisation was linear.} \label{Axicon_Depth}
\end{figure}
\begin{figure}[tb]
\centering \includegraphics[width=1\linewidth]{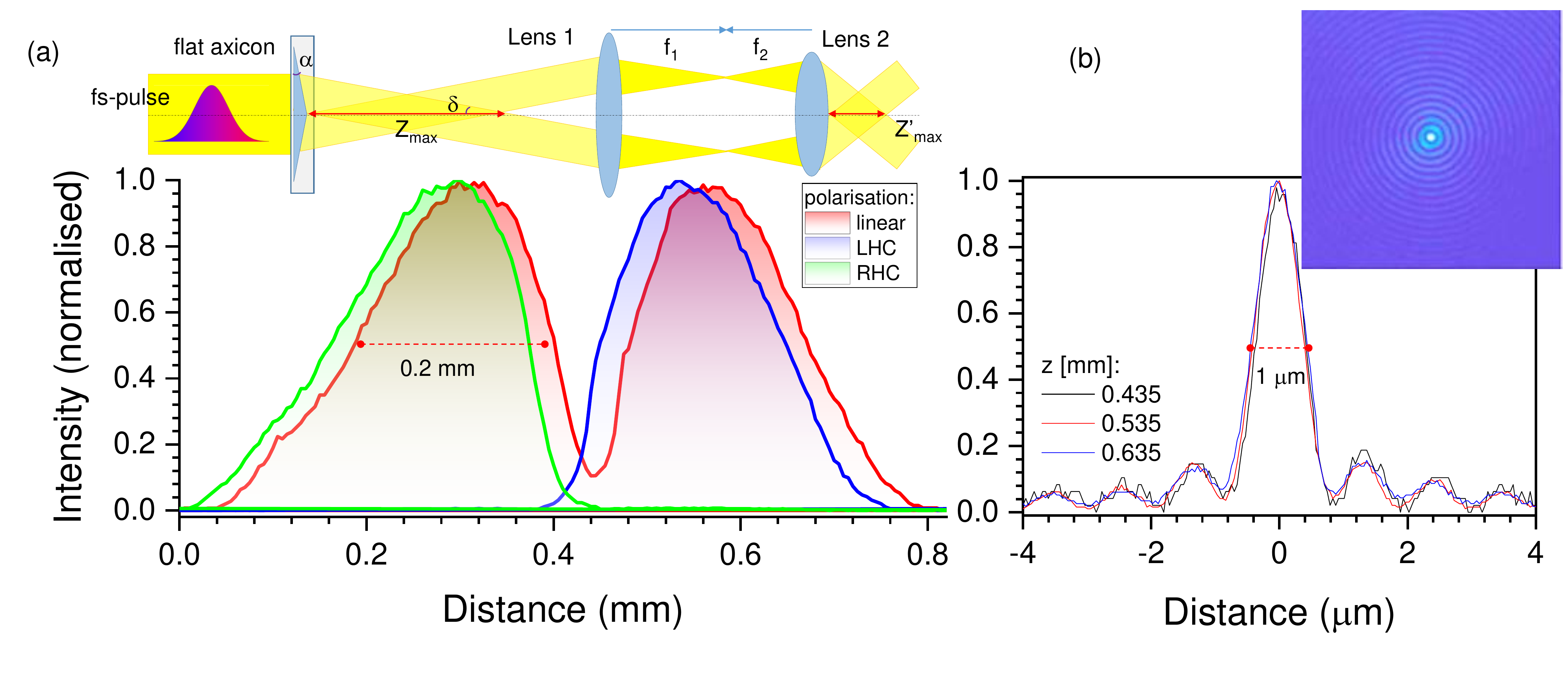}
\caption{(a) Axial intensity distribution of the flat axicon (Fig.~\ref{f-hole}) without a hole for linear and left and right hand polarisations (LHP, RHP) made for wavelength $\lambda = 1030$~nm in silica. Top-inset shows the used 4f-setup to decrease lateral and axial $Z_{max}$  extent of the needle-beam part by factors $\frac{f_1}{f_2}=31.3^\times$ and 976.6$^\times$, respectively. The FWHM of axial intensity cross section was $Z'_{max} = 0.2$~mm, beam quality factor $M^2 - 1.3$, the 4-mm-diameter flat axicon was illuminated with 9.5-mm-beam (at $1/e^2$-intensity). Noteworthy, an on-axis hole in lens-1 and lens-2 would not compromise their focusing performance since the light beam has a doughnut intensity distribution; focal lengths of the lenses-1,2 were $f_1=250$~mm and $f_2=8$~mm. 
(b) Lateral normalized cross section of down-sized Bessel beam at three different axial positions $z$ for LHC polarised illumination; the inset shows a normalised intensity distribution of original Bessel beam on $2.1\times 2.1$~mm$^2$ area.} \label{f-axicon}
\end{figure}

\section{Fabrication of holey-axicon}

Figure~\ref{f-hole} shows an axicon with central 0.1-mm-diameter hole for the optical (pump) and X-ray (probe) using ultra-short 25~fs X-FEL. Collinear (co-axial) propagation geometry facilitates spatial overlap of the laser ablation/modification zone to be probed with on-axis X-FEL 10~keV fs-pulse for fast imaging of phase transitions and material excitation undergoing the dielectric-to-metallic transition with a fast changing Die-Met state of permittivity~\cite{DieMet}.  

The 4-mm-diameter ($r = 2$~mm) lens was inscribed by fs-laser pulses in $t_{ax} = 8$~hours. The central hole was also made by fs-laser damaging of \ce{SiO2} host with subsequent wet etching in basic solution~\cite{01ol277,04apa1549,06apa99}; all together it took 6~hours. Fabrication was carried out on a commercial fs-fabrication station (WOP-Workshop of Photonics), an integrated solution of fs-laser (Pharos, Light Conversion) with linear stages (Aerotech). Polarisation of fs-pulse was set linear with a controlled azimuth $\theta$.

The refractive index change induced in silica is approximately $\Delta n = 1\times 10^{-3}$ due to form-birefringence~\cite{15pr283} and has an axial extent $d$ for retardance of $\pi = \Delta n\times d$ at 530~nm wavelengths (or $\lambda/2$ waveplate), hence $d\approx 133~\mu$m. The entire cylindrical volume of $V_{ax} = \pi r^2 d\approx 1.67\times 10^9~\mu$m$^3$ and the speed of fabrication was $\frac{V_{ax}}{t_{ax}} = 3.48\times 10^6~\mu$m$^3$/min. Considering that approximately only a half of the volume was laser structured the final speed of laser fabrication for the axicon (without central hole) was $v_{fab}\approx 1.7\times 10^6~\mu$m$^3$/min ($1.7\times 10^{-3}$~mm$^3$/min). This high value was achieved because there was a small number (up to 3) of focal position (depth) re-focusings required for axial stitching which can be made perfectly aligned as reported earlier~\cite{13ome1862}. For fs-laser inscription of damaged regions for through-hole etching, it was necessary ten times refocusing along optical axis for the entire 3~mm thickness of silica blank. 

Figure~\ref{f-grat} shows detailed slow-axis orientation map of the form-birefringent nanogratings in silica. The slow axis is perpendicular to the nanograting planes which defined the form birefringence $\Delta n = n_e - n_o < 0$, where $n_e$ is the extraordinary (along the optical axis) and $n_o$ is the ordinary indices, respectively~\cite{Martynas}. The local azimuth of nanogratings was only dependent on the radial position and was independent on the polar orientation (this condition makes an orientation-independent optical element). Various optical elements for manipulation of spin and orbital momenta of light can be designed based on this principle where both propagation and geometrical phases are invoked to control intensity, phase, and polarisation~\cite{Hasman}. The width of the zone with particular azimuth of nanogratings can be flexibly controlled as well as the axial extent of the fs-inscribed gratings. This provides flexibility to engineer axial intensity distribution and its extent which can be understood from discussion on Fig.~\ref{Figure21}. Fs-laser inscription of refractive index modifications can be made for almost loss-less conditions in terms of light absorption and  only light scattering is a cause of some ($< 10\%$) transmission losses at visble 400 - 700~nm window. Recently, the scattering is further reduced in silica glass~\cite{HBS-LSA}. 

The diffraction patterns recorded at distances of 3 cm, 5 cm, 7 cm and 9 cm and the corresponding intensity profiles are shown in Fig,~\ref{Axicon_Depth}. The diameter of the central on-axis intensity was $d=37~\mu$m at $1/e^2$-intensity maximum. Centimeters long Bessel-like beam propagation is due to large apex angle of $178^\circ$.     

\begin{figure}[tb]
\centering \includegraphics[width=.85\linewidth]{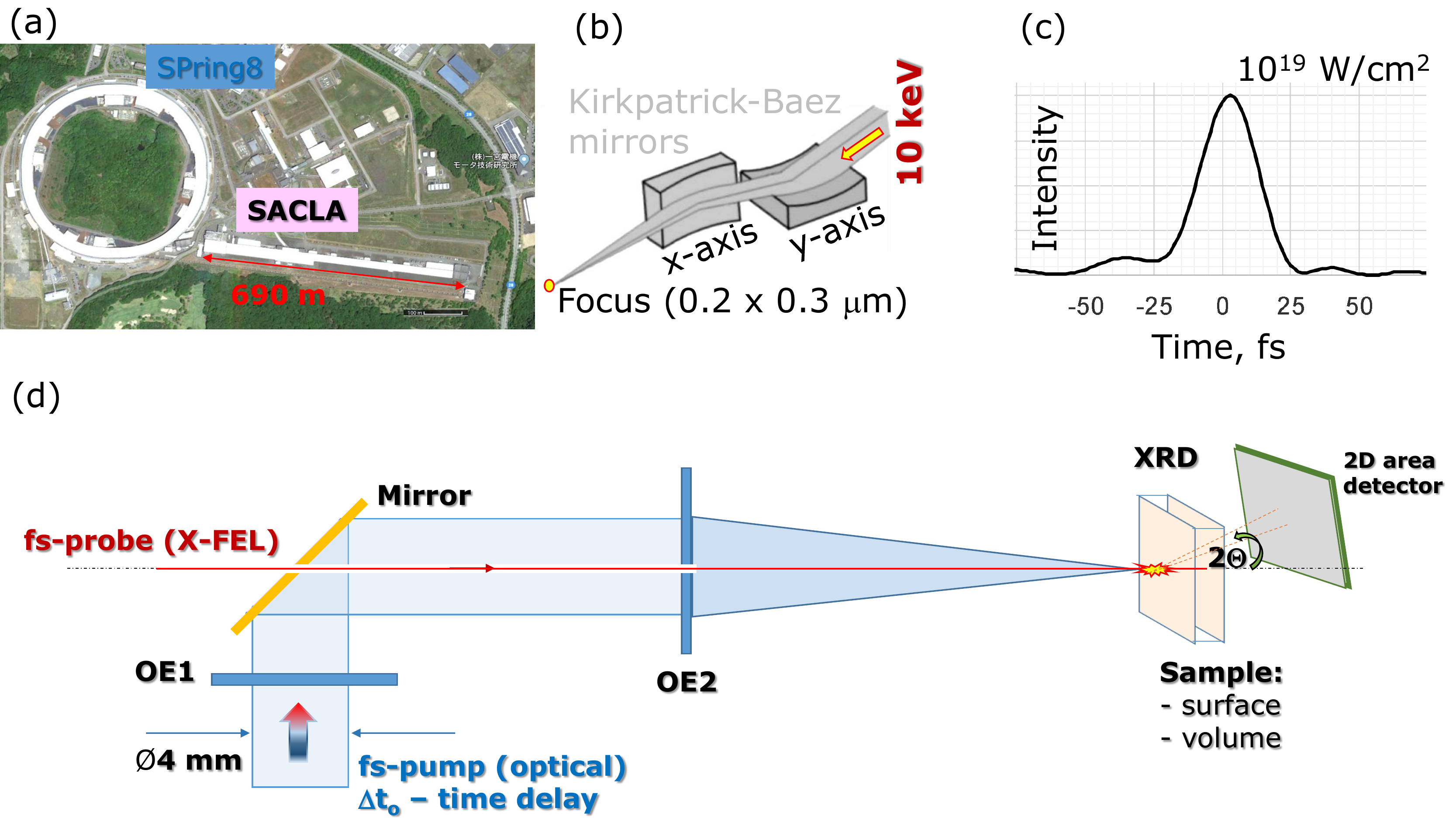}
\caption{Concept of the co-axial volumetric interaction diagnostics (COVID)
using fs-pump (optical) and fs-probe (X-ray). (a) SPring-8 Angstrom Compact free electron LAser (SACLA) facility~\cite{David}. (b) Grazing incidence angle focusing Kirkpatrick-Baez mirrors are used to focus X-FEL beam into typical $200\times 300$~nm$^2$ focal spot; see, X-FELs~\cite{Seniutinas}. (c) X-ray pulse duration $t_x \approx 10$~fs (FWHM) in SACLA reaching $10^{19}$~W/cm$^2$ intensity. (d) Principle of co-axial combination of X-FEL and optical beams using holey-mirror and Bessel beam generator (optical element OE2 or OE1 with 4f-demagnification as shown in the inset of Fig.~\ref{f-axicon}(a)). See several possible implementations of optical pump in Appendix. The optical element OE1 can be used to tailor specific beam intensity distribution~\cite{18jo085606} or used for optical diagnostics of the interaction zone. X-ray diffraction (XRD) measurement is carried out by  X-ray  two-dimensional flat panel detector capable to cover span of 20-90$^\circ$ angular range which is wider than 10$^\circ$ range of a  multi-port  readout  charge-coupled  device  (MPCCD) at 30~cm distance from the target~\cite{mpccd}; the diffraction angle $2\theta$, $\Delta t_o$ is the time delay between the optical and X-ray pulses. } \label{f-xfel}
\end{figure}

Demagnification of the lateral and axial extent of the Bessel beam can be carried out using 4f-optical setup as shown in Fig.~\ref{f-axicon}. The factor of $f_1/f_2\approx 31.3$ was applied resulting in focal volume with approximately $1~\mu$m diameter (Fig.~ \ref{f-axicon}(b)) and axial extent of 0.2~mm at FWHM. The form birefringence $\Delta n$ due to nano-gratings is inscribed in silica with an azimuthally rotating pattern and corresponds to the $\lambda/2$-waveplate condition. This manifests in polarisation insensitivity of the axial intensity distribution (Fig.~ \ref{f-axicon}(a)). There is a difference in leading and trailing slope of the intensity envelope which can be explored in ionisation of the target. Moreover, fs-pulses can be tailored with specific frequency chirp which also contributes to the control of ionisation; red-shifted wavelengths are usually leading the shorter wavelengths (a positive chirp) as schematically shown in the inset of (a). The demonstrated demagnification of Bessel beam down to diameter $2w_0 = 1~\mu$m is favorable to exceed the required irradiance of $I_p\equiv E_p/(t_p\times\pi w_0^2) = 10$~TW/cm$^2$ for dielectric breakdown of dielectrics (sapphire~\cite{06am1361}, olivine~\cite{14aem767}, silica~\cite{11ome605} and glasses~\cite{07njp253}) by only $E_p = 8$~nJ (on target) pulses of $t_p = 100$~fs duration for Gaussian beam and about $E_p\approx 20~\mu$J for the entire length of the Bessel pulse for discussed here holey-axicon. Tight focusing of fs-Bessel pulses into focal spot with lateral cross section $\sim\lambda$ well defines the axial modification inside, e.g., crystalline sapphire~\cite{19n1414}, without formation of nanogratings and confines structural modification directly along optical axis.   

The possibility of controlling the axial intensity envelope by polarisation of the incident light shown for the linear and circular cases can be explored further using elliptically polarised optical pump. Elliptically polarised light is a sum of linearly and circularly polarised contributions providing a well-controlled tunability of the local intensity and its polarisation along the axis in the target.  Another feature of an optical excitation by fs-pump using holey-axicon is presence of $\sim 10\%$ of the pulse energy on optical axis. This creates a condition of pre-excitation of the target since the Bessel part of the beam arrives at an angle $\delta$ with the optical axis. The delay between the side-ray (Bessel beam) and the Gaussian on-axis component can be estimated from the Snell's law $\sin{(\alpha + \delta)} = n_{ax}\sin\alpha$ where $n_{ax} = 1.4$ is the refractive index of axicon (silica), $\alpha = 1^\circ$ is the base angle of axicon. Hence, the length of the hypotenuse is longer than the cathetus (at the angle $\delta$) by $\frac{1}{\cos\delta}-1$. For the conditions shown in Fig.~\ref{f-axicon} with the  beam demagnified by factor $\frac{f_1}{f_2} = 31.3$, the delay of $17\%t_p$ of the pulse length $t_p$ will occur; the angle with axis is increasing as $\delta' = \sin\delta\frac{f_1}{f_2}$. The Bessel-beam component contributes to the resonance absorption mechanism~\cite{18n555} and can create more efficient energy deposition along optical axis. Recent experiment with long ns-pulse Gaussian optical excitation at SACLA showed detailed time evolution of metallic Ta target spalation by XRD monitoring~\cite{Nori}.

The optical pump delivery to the target and co-axial combination of their propagation should be solved considering a specific experimental chamber~\cite{Inubushi}, however it will only be discussed below in terms of the basic concept. In this section we showed the possibility to use fs-laser inscription to fabricate a flat axicon and its performance to reach small lateral diameters (high irradiance) for prolonged axial length of 0.2~mm which elevates tolerances in target-beam alignments. Fs-laser irradiation and wet etching used to make a hole in the flat axicon can be readily used for drilling holes in glass plates used as mirrors for co-linear pump-probe combinations. Micro-holes can also be used to to hold target materials such as polymers, liquid crystals, or liquids for excitation by fs-Bessel pump beam. Metal-coated glass plates with micro-holes are expected to serve as beam alignment aids for both optical and X-FEL beams and can be fabricated via the demonstrated fs-laser fabrication-assisted wet etching.     

\begin{figure}[tb]
\centering \includegraphics[width=0.8\linewidth]{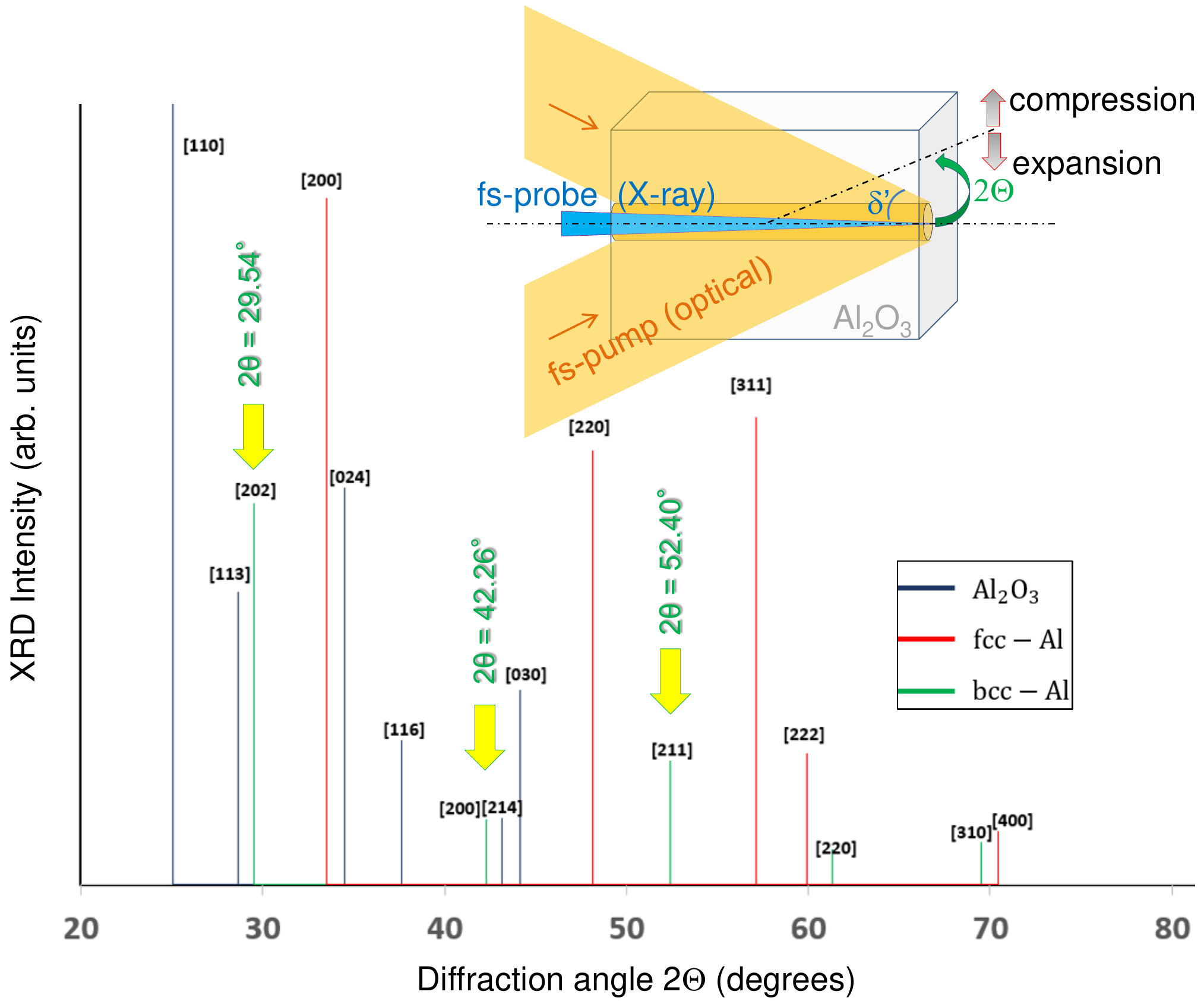}
\caption{XRD simulation for sapphire Al$_2$O$_3$ (ambient conditions) and a common fcc phase of Al (at static pressures between 60 and 100~GPa) and bcc-Al high density phase~\cite{11nc445}. 
Inset shows geometry of experiment using Bessel beam excitation; the diffraction angle $2\theta$ is in respect to the optical axis (horizontal) and the angle $\delta'$ is between the side-rays of Bessel beam with the optical axis (horizontal). Central cylinder shows schematically the core of the Bessel beam where approximately 10\%  of laser pulse energy travels $17\%t_p$ ahead of the main pulse $t_p$ arriving as concentric Bessel maxima onto the optical axis as side-rays (marked by arrows); focusing conditions as in Fig.~\ref{f-axicon}.  } \label{f-xrd}
\end{figure}

\section{Discussion: feasibility of fs-pump(optical) and fs-probe(X-FEL) }\label{con}

Combining optical fs-pump and X-FEL fs-probe is a very promising technique for investigation of light-matter interactions with tens-of-fs temporal resolution and spatial resolution down to few micrometers. Hard X-rays at 10-12~keV at X-FEL (SACLA) do not require a vacuum enclosure for pump-probe experiments which is always a practical advantage for setting up measurements. Noteworthy, vacuum can be used to avoid air breakdown in front of target or at focal points of 4f-optical setup (a negative lens can be used to avoid real focal point with additional advantage for shorter 4f-setup). Targets up to 100~$\mu$m thickness are transparent to the X-FEL (SACLA) beam and interaction of fs-pump(optical) with target in the volume can be monitored by XRD. Co-axial pump-probe alignment is favorable for such experiments due to larger volume of interaction/characterisation. 
Previous studies of micro-void formation inside volume of transparent target were shown after a tightly focused single fs-laser pulse irradiation were structurally and optically investigated (post mortem) after its formation~\cite{06prl166101,06prb214101,11nc445}. \emph{In situ} measurements of void-formation using all-optical fs-pump-probe has a technical limitation due to challenges to focus on the micro-volume of the light-matter interaction zone with two bulky objective lenses at a normal to each other alignment~\cite{17sr10405,11oe5725,11ome1399}; the working distances of objective lenses with numerical aperture $NA\approx 1$ are 0.1-0.3~mm. With a micro-focus of X-FEL beam, probing of the micro-volume of light(optical)-matter interaction at the focal volume has no technical limitation rather a standard micro-precision task of beam alignment.  Figure~\ref{f-xfel} shows SACLA facility~\cite{Ishikawa}, schematics of X-FEL focusing into sub-1~$\mu$m focal spot~\cite{Seniutinas}, temporal profile of the fs-X-ray pulse at 10~keV and the concept of proposed experiment with holey-axicon (Fig.~\ref{f-hole}). The concept drawing has a generic meaning of light delivery onto a target and its actual realisation can be very different adopted to the actual beamline chamber~\cite{Inubushi} and is expected to have 4f-optical setup (inset of Fig.~\ref{f-axicon}(a)) with central holes in lenses. It is noteworthy, that those holes will not deteriorate re-focusing and demagnification of Bessel beam since the light pulse is handled at its doughnut intensity distribution after the axicon. The folding point of optical and X-ray path will be carried out by a mirror with a hole which is a simple fs-fabrication task (Fig.~\ref{f-hole}). Also a reflection from mirror coating onto negative axicon with a central hole can even further simplify co-axial combination of pump and probe.      

The experimental concept outlined in Fig.~\ref{f-xfel}(d) has capability to detect structural modifications using readily available XRD instrumentation developed for SACLA beamlines~\cite{Ishikawa,mpccd} with temporal overlap between 10~fs X-ray probe and optical IR fs-pump solved~\cite{Sato}. Feasibility to increase the volume of shock-wave compressed region in sapphire by approximately 200 times using focused Bessel-beam as compared with a tightly focused Gaussian beam was validated based on existing experimental evidence~\cite{18n555}. High pressure and temperature conditions in an ionised crystalline sapphire would cause separation of Al and O ions with quenching of the affected micro-volume into an amorphous phase with inclusions of bcc-Al nanoparticles tens-of-nm in size~\cite{11nc445}. Figure~\ref{f-xrd} shows the expected angular positions of fcc and bcc-Al~\cite{Polsin,Mitchell} which are in the range of detectable angles at SACLA beamline. 

Surface modification of non-transparent targets can be also investigated with the proposed method. Versatility of this technique can be further augmented with use of additional optical elements (OEs) for shaping pump-beam light intensity distribution or adding extra optical monitoring functions. For example, an OE with dual Fresnel lens plane wave generation can be multiplexed with a conical wave which produces a hologram on the monitoring CCD (added to setup in transmission or reflection). Such a hologram can be used to retrieve the phase and amplitude information from the interaction zone~\cite{20m}. This technique is a single-shot and has 3D spatial and spectral resolution. A toolbox of X-ray optics for X-FEL applications becomes richer with possibility to generate X-ray vortex beams using fs-laser fabricated glass spiral plates~\cite{Nolte}. Also, transient grating induced by X-ray X-FEL and probed by light or X-ray beam was proposed and conceptually tested~\cite{GediminasS}. Shearing interferometry was demonstrated for in situ characterisation of the wavefront of coherent X-FEL beam in the focal region~\cite{GS20}.    

\section{Warm-dense matter (WDM) conditions: 100~eV at solid density plasma}

It was demonstrated during the past decade that ultra-short intense laser pulse tightly focused deep inside a transparent dielectric generates the energy density in excess of several MJ/cm$^3$ (energy per volume is pressure 1~MJ/cm$^3$ = 1~TPa).  Such energy concentration with extremely high heating and quenching rates leads to unusual solid-plasma-solid transformation paths overcoming kinetic barriers to formation of previously unknown high-pressure material phases~\cite{18n555,06prl166101,06prb214101,11nc445,Rapp} some of which can be predicted by ab initio modeling~\cite{Shi}. These discoveries were made with the pulse of Gaussian shape in space and in time.  It was shown recently that the Bessel-shaped pulse transforms much larger amount of a material and allegedly creates higher energy density than was achieved before~\cite{Rapp1,Rapp2}. This makes fs-Bessel pulses promising for creation and characterisation of warm dense matter (WDM) conditions~\cite{wdm1,wdm2}. 

A prominent example of a high pressure material is metallic hydrogen \ce{H2} predicted back in 1935~\cite{h2}. In addition to the profound fundamental interest in the metallic state of this most simple element, it has been suggested that it could be the most powerful rocket fuel yet to exist~\cite{h21} or to exhibit superconductivity at room temperature~\cite{h22}. Predictions suggest that once formed at extreme pressures in excess of 300~GPa ($\sim 370$~GPa is at the Earth's center), it can be recovered metastably to ambient conditions. Thus, it may not only be possible to harness these potentially transformative properties but also to potentially study such metallic \ce{H2} \emph{ex situ} after its creation. Smaller pressures $< 80$~GPa are required to form polymeric solid state \ce{CO2} which could be retrieved to ambient contitions~\cite{PNAS}. 

For research of high-pressure and high-temperature phases it is equally important their retrieval to lower pressure or ambient conditions. Again, ultra-short laser pulses with small volumes are important due to feasibility of ultra-fast thermal quenching which is required to recover metastable material which can also can be stabilised when is in the form of nanocrystals~\cite{11nc445}. Formation of glasses from pure metals is a long-standing goal because they exhibit attractive mechanical properties such as high strength, elasticity and processability. Fundamental understanding of the glass transition is still lacking and is one of the open quests in physics. The very first monoatomic matallic glass was demonstrated when thermal quenching of nano-wires was achieved~\cite{mglass,Dima}. Metastable amorphous \ce{Al2O3} was produced by tightly focused single fs-pulse irradiation of crystalline sapphire~\cite{06am1361}. Even bio-crystal silk can be quenched into its water soluble amorphous form due to small volume and fast quenching~\cite{17mre115028}. Consequence of nonlinear propagation and fillamentation of fs-Bessel pulses and their collapse causing bead-like traces of structural modification in glass~\cite{06ol80} could be revealed with temporal XRD imaging and answer fundamental questions of nonlinear light propagation at strong electronic excitation. Apart from fundamental interest, such understanding would advance applications in laser dicing, drilling, cutting and large volume ablation~\cite{01jjap1197,04jm3358,03apa371,04apa1549}. Research of high-pressure and high temperature phases of materials as well as WDM conditions would benefit basic science and applications from the concept outlined in this memorandum.   
Optical pump with fs-laser Bessel beam combined with synchronised X-ray probe with exceptional optical flux, sub-micron beam size, rather high X-ray photon energy in the keV-range and ultrashort pulse length $\sim$10~fs, allows one, to investigate the atomic motion via X-ray diffraction with extremely high spatial and temporal resolution.  Recently emerging of X-FELs with pulse duration of 10–25~fs and the photon energy $\sim$8~keV currently available at the European X-FEL at DESY in Germany, with up to 17~keV to be expected in near future at the LCLS at the SLAC National Acceleration Laboratory in US, and with 7–15~keV at the SACLA at the RIKEN SPring-8 center in Japan are the ideal sources to be used as a probe pulse to uncover the processes of formation of unusual transient material states in the conditions far from thermodynamic equilibrium.

\section{Outlook for fs-laser produced micro-optics: instrumentation for astronomy}

We showed in the previous discussion that the very same fs-laser material modification is a ready method for fabrication of specific optical elements for tight focusing with added micro-holes for specialised X-FEL optics. The challenge there was to have flexibility for fs-inscription of efficient (low scattering losses and high form birefringence) optical elements which are $\sim$1-mm-diameter and to use the same fs-irradiation for the laser-assisted wet etching of a center-hole in the same design.  

Astrophotonics is an application field of micro-optics where the efficiency of light collection (tight focusing at high transmissivity) is of paramount importance. Large numbers of millimetre-scale lenses are required for applications including multi-object fibre 
spectroscopy, where light is coupled in to hundreds to thousands of single fibres to collect light from many astronomical sources at once over a wide field, and integral field spectroscopy where lens arrays couple light in to integral field units (IFUs) based on image slices or fibre arrays \cite{Parry98}. Multi-object single fibre spectroscopy has a wide range of science applications including the fundamental physics of the Universe \cite{WiggleZ}.  Use of optical fibres allows for less demanding spectrograph construction for wide-fields. Integral field spectroscopy allows detailed spatially resolved measurements of a wide variety of objects; studies of galaxies in particular are a major application \cite{DYNAMO}.

Use of micro-optics allows changes in focal ratio, to optimally feed fibres to reduce focal ratio degradation \cite{PFS} and to allow a contigous focal plane for IFUs \cite{KOALA}. Future instruments \cite{FOBOS} will need a wide-angle (high-NA) aberration-free collection and broad UV-IR spectrum collection capability (i.e. high-performance antireflection coatings) for thousands of micro-lenses. Moreover, solutions for deployment of such instruments in a demanding environment with high thermal fluctuations should be developed. A solution is to integrate all optics into a single block (silica) and use micro-optics. Integration of surface relief lenses, which can be made aspheric for optimised light collection at different wavelengths, with inscribed optical elements as demonstrated in this concept study, provides a promising avenue. Optical fibers can be fixed at the geometrical focus using silica-based resists which have high transparency comparable with fused silica~\cite{10oe10209}. Fiber mounting ports can be etched out for fiber fixation by the very same method shown here for fabrication of the central hole in the axicon lens. Fs-laser assisted cleaving of optical fibers has been demonstrated to deliver smoother surfaces with a median tilt angle approximately 3 times smaller than conventional mechanical fiber cleavers~\cite{Mamun}. This shows that fs-lasers can be used for inscription of optical elements inside the volume of silica or to form surface lenses, to define chemically etched micro-volumes and used for fiber cleaving.

An obvious extension of a single readout port design would consist of a micro-lens array (front) with inscribed optics (middle) and fiber connectors (back). When fixed in an Invar housing, such integration has the benefit of avoiding problems with matching the thermal expansion coefficients of multi-component lenses. Arrays of microlenses on the front surface can be made by fs-laser ablation combined with wet etching and high temperature surface finishing~\cite{Hua,Fan,Cao,Cao1}. Hybrid solutions with polymerised optical elements by fs-laser polymerisation~\cite{16lsa16133} used together with inscribed light collection lenses are also possible, however this deserves a separate and wider discussion. Here we show, that fs-laser fabrication is a ready technology to be utilised in a wide field of applications and can be most effective when the diameter of optics is sub-1-mm (at the current 2020 state-of-the-art).

For a spectrally wide window of operation, optimisation of anti-reflective coatings, lens shape and refractive index profiles will require function specific design (see details in Appendix). The use of form-birefringent nanogratings in silica  allows flexible refractive index engineering with the benefit of low scattering losses up to UV wavelengths~\cite{HBS-LSA}, a spectrally broad functionality, and use of geometrical (Pancharatnam-Berry~\cite{Pancharatnam}; Fig.~\ref{f-grat}) and propagation phase control. The current method of nanograting recording in silica shown in Fig.~\ref{f-grat} caused a small transmissivity decrease to 85\% at 350~nm wavelength after annealing (without anti-reflective coatings). Structural defects induced by nanograting formation in silica can be thermally annealed without change to their form-birefringence~\cite{15pr283} and temperatures up to 1150~$^\circ$C (for 3~hours) were used for silica without erasing optical memory bits~\cite{98jjap1527}. Nanogratings can also be recorded in crystalline sapphire, which has an even broader spectral window of transmission into the IR spectral range as compared to silica, and a very high (metal-like) thermal conductivity~\cite{Fan1}. It was shown that instead of self-organised formation of nanogratings, form-birefringence of $\Delta n\approx 3\times 10^{-3}$ can be inscribed in a deterministic way down to 100~nm periods in sapphire~\cite{Fan1}. Fs-laser fabrication can be used for lens definition on the surface and in the volume of glass (and crystal) as well as laser irradiation assisted wet etching for fabrication of all-in-glass integrated micro-optics and inter-connects with optical fibers.      
\\
\section{Conclusions}

We introduced the concept of a co-axial pump(optical)-probe(X-ray) diagnostic technique based on XRD developed at SACLA. 
It was shown that a fs-laser writing technique for flat optical elements can inscribe the required optical function of an axicon lens (a vortex beam can also be achieved; see Appendix) and can produce a well defined micro-hole for the on-axis combination of optical and X-FEL beams. The estimates provided here show that irradiance in excess of the breakdown of any dielectric ($> 10$~TW/cm$^2$/pulse) are achievable along the axial extension of the Bessel-pulse and therefore lead to structural modifications inside the volume of an optically transparent target. With higher order Bessel beams that have a doughnut intensity distribution in the lateral cross section (see Appendix), it is possible to create an implosion scenario by fs-pump(optical) with X-ray probing of densified material on-axis. The use of ultra-short fs-pulses for both excitation and probing will be able to reveal the intricacies of phase transitions and formation of new materials. The presented discussion is specifically based on the SACLA X-FEL but is applicable to any pump(optical)-probe(X-ray) beamline. The recent demonstration of a narrower $\sim 3$~eV and six times more spectrally bright X-FEL (SACLA) beam by self-seeding of the self-amplified spontaneous emission in the FEL~\cite{new} will increase the sensitivity of detection of structural changes of materials under fs-pump(optical). The concept discussed here is also applicable to other fields, e.g. astronomical instrumentation, where 3D surface lenses can be combined with micro-otical inscribed elements in silica for UV-IR imaging and spectroscopy and all made by fs-laser fabrication.  


\small{\section*{Acknowledgements}
This research was funded by the ARC Discovery DP170100131, DP190103284, Linkage LP190100505 and JST CREST JPMJCR19I3 grants. This work was partially supported by the JSPS Japan-Australia bilateral collaboration program, the MEXT Q-LEAP program (grant no. JPMXS0118070187), and Genesis Research Institute, Inc (Konpon-ken, Toyota). Support of operational costs of Nanotechnology facility by Swinburne University of Technology (in 2016-19) is acknowledged. We are grateful to the Australian Institute of Nuclear Science and Engineering (AINSE) for facilitation of this study. Discussions of pump-probe experiments with Dr. Takeshi Matsuoka are gratefully acknowledged.  
}

\section*{References}
\bibliographystyle{spiebib}
\small\bibliography{paper6c,nanolab1}

\begin{thebibliography}{100}

\bibitem{David}
D.~Pile, ``First light from {SACLA},'' {\em Nat. Photonics}~{\bf 5}(8),
  p.~456–457, 2011.

\bibitem{01apl725}
T.~Kondo, S.~Matsuo, S.~Juodkazis, and H.~Misawa, ``A novel femtosecond laser
  interference technique with diffractive beam splitter for fabrication of
  three-dimensional photonic crystals,'' {\em Appl. Phys. Lett..}~{\bf 79}(6),
  pp.~725--727, 2001.

\bibitem{Sato15}
T.~Sato, T.~Togashi, K.~Ogawa, T.~Katayama, Y.~Inubushi, K.~Tono, and
  M.~Yabashi, ``Highly efficient arrival timing diagnostics for femtosecond
  {X}-ray and optical laser pulses,'' {\em Appl. Phys. Express}~{\bf 8},
  p.~012702, 2015.

\bibitem{fel}
B.~McNeil and N.~R. Thompson, ``X-ray free-electron lasers,'' {\em Nat.
  Photonics}~{\bf 4}(12), p.~814–821, 2010.

\bibitem{Pikuz}
P.~Mabey, B.~Albertazzi, T.~Michel, G.~Rigon, S.~Makarov, N.~Ozaki,
  T.~Matsuoka, S.~Pikuz, T.~Pikuz, and M.~Koenig, ``Characterization of high
  spatial resolution lithium fluoride {X-ray} detectors,'' {\em Rev. Sci.
  Instrum.}~{\bf 90}, p.~063702, 2019.

\bibitem{01ol277}
A.~Marcinkevicius, S.~Juodkazis, M.~Watanabe, M.~Miwa, S.~Matsuo, H.~Misawa,
  and J.~Nishii, ``Femtosecond laser-assisted three-dimensional
  microfabrication in silica,'' {\em Opt. Lett.}~{\bf 26}(5), pp.~277--279,
  2001.

\bibitem{04apa1549}
S.~Juodkazis, K.~Yamasaki, V.~Mizeikis, S.~Matsuo, and H.~Misawa, ``Formation
  of embedded patterns in glasses using femtosecond irradiation,'' {\em Appl.
  Phys. A}~{\bf 79}(4-6), pp.~1549 -- 1553, 2004.

\bibitem{06apa99}
S.~Matsuo, Y.~Tabuchi, T.~Okada, S.~Juodkazis, and H.~Misawa, ``Femtosecond
  laser assisted etching of quartz: microstructuring from inside,'' {\em Appl.
  Phys. A.}~{\bf 84}(1 -- 2), pp.~99 -- 102, 2007.

\bibitem{Inubushi}
Y.~Inubushi, T.~Yabuuchi, T.~Togashi, K.~Sueda, K.~Miyanishi, Y.~Tange,
  N.~Ozaki, T.~Matsuoka, R.~Kodama, T.~Osaka, S.~Matsuyama, K.~Yamauchi,
  H.~Yumoto, T.~Koyama, H.~Ohashi, K.~Tono, and M.~Yabashi, ``Development of an
  experimental platform for combinative use of an {XFEL} and a high-power
  nanosecond laser,'' {\em Appl. Sci.}~{\bf 10}, p.~2224, 2020.

\bibitem{wdm1}
S.~Chang, ``Extreme heating with an x-ray free-electron laser,'' {\em Physics
  Today}~{\bf 68}(5), p.~18, 2015.

\bibitem{wdm2}
P.~Drake, ``High-energy-density physics,'' {\em Physics Today}~{\bf 63}(6),
  p.~28, 2010.

\bibitem{Golub}
I.~Golub, ``Fresnel axicon,'' {\em Opt. Lett.}~{\bf 31}, pp.~1890--1892, 2006.

\bibitem{Vijayakumar}
G.~Gervinskas, G.~Seniutinas, A.~Vijayakumar, S.~Bhattacharya, E.~Jelmakas,
  A.~Kadys, R.~Toma\v{s}i\={u}nas, and S.~Juodkazis, ``Fabrication and
  replication of micro-optical structures for growth of {GaN}-based light
  emitting diodes,'' in {\em Proc. SPIE 8923, Micro+Nano Materials, Devices and
  Applications},  J.~Friend, ed., p.~89234L, 2013.

\bibitem{McLeod}
J.~McLeod, ``The axicon: A new type of optical element,'' {\em J. Opt. Soc.
  Am.}~{\bf 44}, pp.~592--597, 1954.

\bibitem{Scott}
G.~Scott and N.~McArdle, ``Efficient generation of nearly diffraction-free
  beams using an axicon,'' {\em Opt. Eng.}~{\bf 31}, p.~2640–2643, 1992.

\bibitem{Kizuka}
Y.~Kizuka, M.~Yamauchi, and Y.~Matsuoka, ``Characteristics of a laser beam spot
  focused by a binary diffractive axicon,'' {\em Opt. Eng.}~{\bf 471},
  p.~053401, 2008.

\bibitem{Bouchal}
Z.~Bouchal, J.~Wagner, and M.~Chlup, ``Self-reconstruction of a distorted
  nondiffracting beam,'' {\em Opt. Commun.}~{\bf 151}, p.~207–211, 1998.

\bibitem{VijayakumarSPIE}
A.~Vijayakumar and S.~Bhattacharya, {\em Design and Fabrication of Diffractive
  Optical Elements with {MATLAB}}, ch.~1.
\newblock SPIE, Bellingham WA, USA, 2017.

\bibitem{DieMet}
E.~Gamaly and A.~Rode, ``Ultrafast re-structuring of the electronic landscape
  of transparent dielectrics: new material states {(Die-Met)},'' {\em Appl.
  Phys. A}~{\bf 124}(3), p.~278, 2018.

\bibitem{15pr283}
C.~J. de~Jong, A.~Lajevardipour, M.~Gecevicius, M.~Beresna, G.~Gervinskas,
  P.~G. Kazansky, Y.~Bellouard, A.~H.~A. Clayton, and S.~Juodkazis, ``{Deep-UV}
  fluorescence lifetime imaging microscopy,'' {\em Photon. Research}~{\bf
  3}(5), pp.~283 -- 288, 2015.

\bibitem{13ome1862}
M.~Mikutis, T.~Kudrius, G.~\v{S}lekys, D.~Paipulas, and S.~Juodkazis, ``High
  90{\%} efficiency {Bragg} gratings formed in fused silica by femtosecond
  {Gauss-Bessel} laser beams,'' {\em Opt. Mat. Express}~{\bf 3}(11),
  pp.~1862--1871, 2013.

\bibitem{Martynas}
M.~B.~M. Gecevi\v{c}ius, P.~Kazansky, and T.~Gertus, ``Radially polarized
  optical vortex converter created by femtosecond laser nanostructuring of
  glass,'' {\em Appl. Phys. Lett.}~{\bf 98}(20), p.~201101, 2011.

\bibitem{Hasman}
D.~Lin, P.~Fan, E.~Hasman, and M.~L. Brongersma, ``Dielectric gradient
  metasurface optical elements,'' {\em Science}~{\bf 345}(6194), pp.~298--302,
  2014.

\bibitem{HBS-LSA}
M.~Sakakura, Y.~Lei, L.~Wang, Y.-H. Yu, and P.~G. Kazansky, ``Ultralow-loss
  geometric phase and polarization shaping by ultrafast laser writing in silica
  glass,'' {\em Light Sci. Appl.}~{\bf 9}, p.~15, 2020.

\bibitem{Seniutinas}
M.~Makita, G.~Seniutinas, M.~Seaberg, H.~Lee, E.~Galtier, M.~Liang, A.~Aquila,
  S.~Boutet, A.~Hashim, M.~S. Hunter, T.~{van Driel}, U.~Zastrau, C.~David, and
  B.~Nagler, ``Double grating shearing interferometry for {X-ray} free-electron
  laser beams,'' {\em Optica}~{\bf 7}(5), pp.~404--409, 2020.

\bibitem{18jo085606}
Dharmavarapu, S.~Bhattacharya, and S.~Juodkazis, ``Diffractive optics for axial
  intensity shaping of {Bessel} beams,'' {\em J. Opt.}~{\bf 20}, p.~085606,
  2018.

\bibitem{mpccd}
T.~Kameshima, S.~Ono, T.~Kudo, K.~Ozaki, Y.~Kirihara, K.~Kobayashi,
  Y.~Inubushi, M.~Yabashi, T.~Horigome, A.~Holland, K.~Holland, D.~Burt,
  H.~Murao, and T.~Hatsui, ``Development of an {X-ray} pixel detector with
  multi-port charge-coupled device for {X-ray} free-electron laser
  experiments,'' {\em Rev. Sci. Instrum.}~{\bf 85}, p.~033110, 2014.

\bibitem{06am1361}
S.~Juodkazis, K.~Nishimura, H.~Misawa, T.~Ebisui, R.~Waki, S.~Matsuo, and
  T.~Okada, ``Control over the state of crystallinity: Sapphire,'' {\em Adv.
  Mat.}~{\bf 18}(11), pp.~1361 -- 1364, 2006.

\bibitem{14aem767}
R.~Buividas, G.~Gervinskas, A.~Tadich, B.~C.~C. Cowie, V.~Mizeikis,
  A.~Vailionis, D.~de~Ligny, E.~G. Gamaly, A.~V. Rode, and S.~Juodkazis,
  ``Phase transformation in laser-induced micro-explosion in olivine
  {(Fe,Mg)$_2$SiO$_4$},'' {\em Adv. Eng. Mat.}~{\bf 16}(6), pp.~767 -- 773,
  2014.

\bibitem{11ome605}
L.~Bressel, D.~de~Ligny, C.~Sonneville, V.~Martinez, V.~Mizeikis, R.~Buividas,
  and S.~Juodkazis, ``Femtosecond laser induced density changes in {GeO$_2$}
  and {SiO$_2$} glasses: fictive temperature effect,'' {\em Opt. Mat.
  Express}~{\bf 1}(4), pp.~605--613, 2011.

\bibitem{07njp253}
T.~Hashimoto, S.~Juodkazis, and H.~Misawa, ``Void formation in glass,'' {\em
  New. J. Phys.}~{\bf 9}, pp.~253 /1--9, 2007.

\bibitem{19n1414}
H.~Fan, M.~Ryu, R.~Honda, J.~Morikawa, Z.-Z. Li, L.~Wang, J.~Maksimovic,
  S.~Juodkazis, Q.-D. Chen, and H.-B. Sun, ``Laser-inscribed stress-induced
  birefringence of sapphire,'' {\em Nanomaterials}~{\bf 9}(10), p.~1414, 2019.

\bibitem{18n555}
E.~G. Gamaly, S.~Juodkazis, and A.~V. Rode, ``Extreme energy density confined
  inside a transparent crystal: Status and perspectives of solid-plasma-solid
  transformations,'' {\em Nanomaterials}~{\bf 8}(7), p.~555, 2018.

\bibitem{Nori}
B.~Albertazzi, N.~Ozaki, V.~Zhakhovsky, A.~Faenov, H.~Habara, M.~Harmand, N.~J.
  Hartley, D.~Ilnitsky, N.~Inogamov, Y.~Inubushi, T.~Ishikawa, T.~Katayama,
  M.~Koenig, A.~Krygier, T.~Matsuoka, S.~Matsuyama, E.~McBride, K.~Migdal,
  G.~Morard, T.~Okuchi, T.~Pikuz, O.~Sakata, Y.~Sano, T.~Sato, T.~Sekine,
  T.~Seto, K.~Takahashi, H.~Tanaka, K.~A. Tanaka, Y.~Tange, T.~Togashi,
  K.~Tono, Y.~Umeda, T.~Vinci, M.~Yabashi, T.~Yabuuchi, K.~Yamauchi, and
  R.~Kodama, ``Dynamic fracture of tantalum under extreme tensile stress,''
  {\em Sci. Adv.}~{\bf 3}, p.~e1602705, 2017.

\bibitem{11nc445}
A.~Vailionis, E.~G. Gamaly, V.~Mizeikis, W.~Yang, A.~Rode, and S.~Juodkazis,
  ``Evidence of super-dense {Aluminum} synthesized by ultra-fast
  micro-explosion,'' {\em Nature Communications}~{\bf 2}, p.~445, 2011.

\bibitem{06prl166101}
S.~Juodkazis, K.~Nishimura, S.~Tanaka, H.~Misawa, E.~E. Gamaly,
  B.~Luther-Davies, L.~Hallo, P.~Nicolai, and V.~Tikhonchuk, ``Laser-induced
  microexplosion confined in the bulk of a sapphire crystal: Evidence of
  multimegabar pressures,'' {\em Phys. Rev. Lett.}~{\bf 96}(16), p.~166101,
  2006.

\bibitem{06prb214101}
E.~E. Gamaly, S.~Juodkazis, K.~Nishimura, H.~Misawa, B.~Luther-Davies,
  L.~Hallo, P.~Nicolai, and V.~Tikhonchuk, ``Laser-matter interaction in a bulk
  of a transparent solid: confined micro-explosion and void formation,'' {\em
  Phys. Rev. B}~{\bf 73}, p.~214101, 2006.

\bibitem{17sr10405}
T.~Hayasaki, S.-I. Fukuda, S.~Hasegawa, and S.~Juodkazis, ``Two-color
  pump-probe interferometry of ultra-fast light-matter interaction,'' {\em Sci.
  Reports}~{\bf 7}, p.~10405, 2017.

\bibitem{11oe5725}
Y.~Hayasaki, M.~Isaka, A.~Takita, and S.~Juodkazis, ``Time-resolved
  interferometry of femtosecond-laser induced processes under tight focusing
  and close-to optical breakdown inside borosilicate glass,'' {\em Opt.
  Express}~{\bf 19}(7), pp.~5725--5734, 2011.

\bibitem{11ome1399}
Y.~Hayasaki, K.~Iwata, S.~Hasegawa, A.~Takita, and S.~Juodkazis,
  ``Time-resolved axial-view of the dielectric breakdown under tight focusing
  in glass,'' {\em Opt. Mater. Express}~{\bf 1}, pp.~1399--1408, 2011.

\bibitem{Ishikawa}
M.~Yabashi, H.~Tanaka, and T.~Ishikawa, ``Overview of the {SACLA} facility,''
  {\em J. Synchr. Rad.}~{\bf 22}(3), pp.~477--484, 2015.

\bibitem{Sato}
T.~Sato, T.~Togashi, K.~Ogawa, T.~Katayama, Y.~Inubushi, K.~Tono, and
  M.~Yabashi, ``Highly efficient arrival timing diagnostics for femtosecond
  {X-ray} and optical laser pulses,'' {\em Appl. Phys. Express}~{\bf 8},
  p.~012702, 2015.

\bibitem{Polsin}
D.~N. Polsin, D.~E. Fratanduono, J.~R. Rygg, A.~Lazicki, R.~F. Smith, J.~H.
  Eggert, M.~C. Gregor, B.~H. Henderson, J.~A. Delettrez, R.~G. Kraus, P.~M.
  Celliers, F.~Coppari, D.~C. Swift, C.~A. McCoy, C.~T. Seagle, J.-P. Davis,
  S.~J. Burns, G.~W. Collins, and T.~R. Boehl, ``Measurement of
  body-centered-cubic aluminum at {475 GPa},'' {\em Phys. Rev. Lett.}~{\bf
  120}, p.~029902, 2018.

\bibitem{Mitchell}
A.~C. Mitchell, W.~J. Nellis, J.~A. Moriarty, R.~A. Heinle, N.~C. Holmes, R.~E.
  Tipton, and G.~W. Repp, ``Equation of state of {Al, Cu, Mo, and Pb} at shock
  pressures up to {2.4 TPa (24 Mbar)},'' {\em J. Appl. Phys.}~{\bf 69},
  p.~2981, 1991.

\bibitem{20m}
V.~Anand, T.~Katkus, and S.~Juodkazis, ``Randomly multiplexed diffractive lens
  and axicon for spatial and spectral imaging,'' {\em Micromachines}~{\bf
  11}(4), p.~437, 2020.

\bibitem{Nolte}
F.~Seiboth, M.~Kahnt, M.~Lyubomirskiy, M.~Seyrich, F.~Wittwer, T.~Ullsperger,
  S.~Nolte, D.~Batey, C.~Rau, and C.~G. Schroer, ``Refractive hard x-ray vortex
  phase plates,'' {\em Opt. Lett.}~{\bf 44}(18), pp.~4622--4625, 2019.

\bibitem{GediminasS}
C.~Svetina, R.~Mankowsky, G.~Knopp, F.~Koch, G.~Seniutinas, B.~R\"{o}sner,
  A.~Kubec, M.~Lebugle, I.~Mochi, M.~Beck, C.~Cirelli, J.~Krempasky,
  C.~Pradervand, J.~Rouxel, G.~F. Mancini, S.~Zerdane, B.~Pedrini, V.~Esposito,
  G.~Ingold, U.~Wagner, U.~Flechsig, R.~Follath, M.~Chergui, C.~Milne, H.~T.
  Lemke, C.~David, and P.~Beaud, ``Refractive hard x-ray vortex phase plates,''
  {\em Opt. Lett.}~{\bf 44}(3), pp.~574--577, 2019.

\bibitem{GS20}
M.~Makita, G.~Seniutinas, M.~Seaberg, H.~Lee, E.~Galtier, M.~Liang, A.~Aquila,
  S.~Boutet, A.~Hashim, M.~Hunter, T.~van Driel, U.~Zastrau, C.~David, and
  B.~Nagler, ``Double grating shearing interferometry for x-ray free-electron
  laser beams,'' {\em Optica}~{\bf 7}(5), pp.~404--409, 2020.

\bibitem{Rapp}
L.~Rapp, B.~Haberl, C.~J. Pickard, J.~E. Bradby, E.~G. Gamaly, J.~S. Williams,
  and A.~V. Rode, ``Experimental evidence of new tetragonal polymorphs of
  silicon formed through ultrafast laser-induced confined microexplosion,''
  {\em Nat. Comm.}~{\bf 6}, p.~7555, 2015.

\bibitem{Shi}
X.~Shi, C.~He, C.~J. Pickard, C.~Tang, and J.~Zhong, ``Stochastic generation of
  complex crystal structures combining group and graph theory with application
  to carbon,'' {\em Phys. Rev. B}~{\bf 97}, p.~014104, 2018.

\bibitem{Rapp1}
L.~Rapp, R.~Meyer, R.~Giust, L.~Furfaro, M.~Jacquot, P.~A. Lacourt, J.~M.
  Dudley, and F.~Courvoisier, ``High aspect ratio micro-explosions in the bulk
  of sapphire generated by femtosecond {Bessel} beams,'' {\em Sci.
  Reports}~{\bf 6}, p.~34286, 2016.

\bibitem{Rapp2}
E.~G. Gamaly, A.~V. Rode, L.~Rapp, R.~Giust, L.~Furfaro, P.~A. Lacourt, J.~M.
  Dudley, F.~Courvoisier, and S.~Juodkazis, ``Interaction of the ultra-short
  bessel beam with transparent dielectrics: Evidence of high-energy
  concentration and multi-tpa pressure,'' {\em arXiv:1708.07630} , 2017.

\bibitem{h2}
M.~Eremets and I.~Troyan, ``Conductive dense hydrogen,'' {\em Nature
  Materials}~{\bf 10}, pp.~927--931, 2011.

\bibitem{h21}
I.~Silvera and J.~Cole, ``Metallic hydrogen: The most powerful rocket fuel yet
  to exist,'' {\em J. Phys. Conf. Ser.}~{\bf 215}, p.~012194, 2010.

\bibitem{h22}
N.~Ashcroft, ``Hydrogen dominant metallic alloys: High-temperature
  semiconductors?,'' {\em Phys. Rev. Lett.}~{\bf 92}, p.~187002, 2004.

\bibitem{PNAS}
J.~Sun, D.~D. Klug, R.~Marto\v{n}\'{a}k, J.~A. Montoya, M.-S. Lee, S.~Scandolo,
  and E.~Tosatti, ``High-pressure polymeric phases of carbon dioxide,'' {\em
  PNAS}~{\bf 106}(15), pp.~6077--6081, 2009.

\bibitem{mglass}
L.~Zhong, J.~Wang, H.~Sheng, Z.~Zhang, and S.~X. Mao, ``Formation of monoatomic
  metallic glasses through ultrafast liquid quenching,'' {\em Nature}~{\bf
  512}, pp.~177--182, 2014.

\bibitem{Dima}
D.-M. Tang, C.-L. Ren, R.~Lv, W.-J. Yu, P.-X. Hou, M.-S. Wang, X.~Wei, Z.~Xu,
  N.~Kawamoto, Y.~Bando, M.~Mitome, C.~Liu, H.-M. Cheng, and D.~Golberg,
  ``Amorphization and directional crystallization of metals confined in carbon
  nanotubes investigated by in situ transmission electron microscopy,'' {\em
  Nano Lett.}~{\bf 15}(8), pp.~4922--4927, 2015.

\bibitem{17mre115028}
M.~Ryu, H.~Kobayashi, A.~Balcytis, X.~Wang, J.~Vongsvivut, J.~Li, N.~Urayama,
  V.~Mizeikis, M.~Tobin, and S.~Juodkazis, ``Nanoscale chemical mapping of
  laser-solubilized silk,'' {\em Mater. Res. Express}~{\bf 4}, p.~115028, 2017.

\bibitem{06ol80}
E.~Gai\v{z}auskas, E.~Vanagas, V.~Jarutis, S.~Juodkazis, V.~Mizeikis, and
  H.~Misawa, ``Discrete damage traces from filamentation of {B}essel-{G}auss
  pulses,'' {\em Opt. Lett.}~{\bf 31}(1), pp.~80--82, 2006.

\bibitem{01jjap1197}
A.~Marcinkevicius, S.~Juodkazis, S.~Matsuo, V.~Mizeikis, and H.~Misawa,
  ``Application of \textrm{Bessel} beams for microfabrication of dielectrics by
  femtosecond laser,'' {\em Jpn. J. Appl. Phys.}~{\bf 40}(11A),
  pp.~L1197--L1199, 2001.

\bibitem{04jm3358}
E.~Vanagas, J.~Kawai, D.~Tuzilin, I.~Kudryashov, A.~Mizuyama, K.~G. Nakamura,
  K.-I. Kondo, S.-Y. Koshihara, M.~Takesada, K.~Matsuda, S.~Juodkazis,
  V.~Jarutis, S.~Matsuo, and H.~Misawa, ``Glass cutting by femtosecond pulsed
  irradiation,'' {\em J. Microlith. Microfab. Microsyst.}~{\bf 3}(2),
  pp.~358--363, 2004.

\bibitem{03apa371}
K.~Yamasaki, S.~Juodkazis, S.~Matsuo, and H.~Misawa, ``Three-dimensional
  microchannels in polymers: one step fabrication,'' {\em Appl. Phys. A}~{\bf
  77}, pp.~371--373, 2003.

\bibitem{Parry98}
I.~{Parry}, ``{The Astronomical Uses of Optical Fibers},'' in {\em Fiber Optics
  in Astronomy III},  S.~{Arribas}, E.~{Mediavilla}, and F.~{Watson}, eds.,
  {\em Astronomical Society of the Pacific Conference Series} {\bf 152}, p.~3,
  Jan. 1998.

\bibitem{WiggleZ}
D.~{Parkinson}, S.~{Riemer-S{\o}rensen}, C.~{Blake}, G.~B. {Poole}, T.~M.
  {Davis}, S.~{Brough}, M.~{Colless}, C.~{Contreras}, W.~{Couch}, S.~{Croom},
  D.~{Croton}, M.~J. {Drinkwater}, K.~{Forster}, D.~{Gilbank}, M.~{Gladders},
  K.~{Glazebrook}, B.~{Jelliffe}, R.~J. {Jurek}, I.~h. {Li}, B.~{Madore}, D.~C.
  {Martin}, K.~{Pimbblet}, M.~{Pracy}, R.~{Sharp}, E.~{Wisnioski}, D.~{Woods},
  T.~K. {Wyder}, and H.~K.~C. {Yee}, ``{The WiggleZ Dark Energy Survey: Final
  data release and cosmological results},'' {\em Phys. Rev. D.}~{\bf 86},
  p.~103518, Nov. 2012.

\bibitem{DYNAMO}
A.~W. {Green}, K.~{Glazebrook}, P.~J. {McGregor}, R.~G. {Abraham}, G.~B.
  {Poole}, I.~{Damjanov}, P.~J. {McCarthy}, M.~{Colless}, and R.~G. {Sharp},
  ``{High star formation rates as the origin of turbulence in early and modern
  disk galaxies},'' {\em Nature}~{\bf 467}, pp.~684--686, Oct. 2010.

\bibitem{PFS}
N.~{Tamura}, N.~{Takato}, A.~{Shimono}, Y.~{Moritani}, K.~{Yabe},
  Y.~{Ishizuka}, A.~{Ueda}, Y.~{Kamata}, H.~{Aghazarian}, S.~{Arnouts},
  G.~{Barban}, R.~H. {Barkhouser}, R.~C. {Borges}, D.~F. {Braun}, M.~A. {Carr},
  P.-Y. {Chabaud}, Y.-C. {Chang}, H.-Y. {Chen}, M.~{Chiba}, R.~C.~Y. {Chou},
  Y.-H. {Chu}, J.~{Cohen}, R.~P. {de Almeida}, A.~C. {de Oliveira}, L.~S. {de
  Oliveira}, R.~G. {Dekany}, K.~{Dohlen}, J.~B. {dos Santos}, L.~H. {dos
  Santos}, R.~{Ellis}, M.~{Fabricius}, D.~{Ferrand}, D.~{Ferreira},
  M.~{Golebiowski}, J.~E. {Greene}, J.~{Gross}, J.~E. {Gunn}, R.~{Hammond},
  A.~{Harding}, M.~{Hart}, T.~M. {Heckman}, C.~M. {Hirata}, P.~{Ho}, S.~C.
  {Hope}, L.~{Hovland}, S.-F. {Hsu}, Y.-S. {Hu}, P.-J. {Huang}, M.~{Jaquet},
  Y.~{Jing}, J.~{Karr}, M.~{Kimura}, M.~E. {King}, E.~{Komatsu}, V.~{Le Brun},
  O.~{Le F{\`e}vre}, A.~{Le Fur}, D.~{Le Mignant}, H.-H. {Ling}, C.~P.
  {Loomis}, R.~H. {Lupton}, F.~{Madec}, P.~{Mao}, L.~S. {Marrara}, C.~{Mendes
  de Oliveira}, Y.~{Minowa}, C.~{Morantz}, H.~{Murayama}, G.~J. {Murray},
  Y.~{Ohyama}, J.~{Orndorff}, S.~{Pascal}, J.~M. {Pereira}, D.~{Reiley},
  M.~{Reinecke}, A.~{Ritter}, M.~{Roberts}, M.~A. {Schwochert}, M.~D.
  {Seiffert}, S.~A. {Smee}, L.~{Sodre}, D.~N. {Spergel}, A.~J. {Steinkraus},
  M.~A. {Strauss}, C.~{Surace}, Y.~{Suto}, N.~{Suzuki}, J.~{Swinbank}, P.~J.
  {Tait}, M.~{Takada}, T.~{Tamura}, Y.~{Tanaka}, L.~{Tresse}, O.~{Verducci},
  D.~{Vibert}, C.~{Vidal}, S.-Y. {Wang}, C.-Y. {Wen}, C.-H. {Yan}, and
  N.~{Yasuda}, ``{Prime Focus Spectrograph (PFS) for the Subaru telescope:
  overview, recent progress, and future perspectives},'' in {\em Proc. SPIE},
  {\em Society of Photo-Optical Instrumentation Engineers (SPIE) Conference
  Series} {\bf 9908}, p.~99081M, Aug. 2016.

\bibitem{KOALA}
R.~{Zhelem}, J.~{Brzeski}, S.~{Case}, V.~{Churilov}, S.~{Ellis}, T.~{Farrell},
  A.~{Green}, A.~{Heng}, A.~{Horton}, M.~{Ireland }, D.~{Jones}, U.~{Klauser},
  J.~{Lawrence}, S.~{Miziarski}, D.~{Orr}, N.~{Pai}, N.~{Staszak}, J.~{Tims},
  M.~{Vuong}, L.~{Waller}, and P.~{Xavier}, ``{KOALA, a wide-field 1000 element
  integral-field unit for the Anglo-Australian Telescope: assembly and
  commissioning},'' in {\em Proc. SPIE},  {\em Society of Photo-Optical
  Instrumentation Engineers (SPIE) Conference Series} {\bf 9147}, p.~91473K,
  July 2014.

\bibitem{FOBOS}
K.~{Bundy}, K.~{Westfall}, N.~{MacDonald}, R.~{Kupke}, M.~{Savage},
  C.~{Poppett}, A.~{Alabi}, G.~{Becker}, J.~{Burchett}, P.~{Capak}, A.~{Coil},
  M.~{Cooper}, D.~{Cowley}, W.~{Deich}, D.~{Dillon}, J.~{Edelstein},
  P.~{Guhathakurta}, J.~{Hennawi}, M.~{Kassis}, K.~G. {Lee}, D.~{Masters},
  T.~{Miller}, J.~{Newman}, J.~{O'Meara}, J.~X. {Prochaska}, M.~{Rau},
  J.~{Rhodes}, R.~M. {Rich}, C.~{Rockosi}, A.~{Romanowsky}, C.~{Schafer},
  D.~{Schlegel}, A.~{Shapley}, B.~{Siana}, Y.~S. {Ting}, D.~{Weisz},
  M.~{White}, B.~{Williams}, G.~{Wilson}, M.~{Wilson}, and R.~{Yan}, ``{FOBOS:
  A Next-Generation Spectroscopic Facility},'' in {\em Astro2020: Decadal
  Survey on Astronomy and Astrophysics, APC white papers, Bulletin of the
  American Astronomical Society},   {\bf 51}, p.~198, Sept. 2019.

\bibitem{10oe10209}
M.~Malinauskas, A.~\v{Z}ukauskas, G.~Bi\v{c}kauskait\.{e}, R.~Gadonas, and
  S.~Juodkazis, ``Mechanisms of three-dimensional structuring of photo-polymers
  by tightly focussed femtosecond laser pulses,'' {\em Opt. Express}~{\bf
  18}(10), pp.~10209--10221, 2010.

\bibitem{Mamun}
M.~{Al Mamun}, P.~J. Cadusch, T.~Katkus, S.~Juodkazis, and P.~R. Stoddart,
  ``Quantification of fiber end-face quality: femtosecond laser versus
  mechanical cleaving,'' {\em Opt. Express} , p.~(submitted), 2020.

\bibitem{Hua}
J.-G. Hua, H.~Ren, A.~Jia, Z.-N. Tian, L.~Wang, S.~Juodkazis, Q.-D. Chen, and
  H.-B. Sun, ``Convex silica microlens arrays via femtosecond laser writing,''
  {\em Optics Lett.}~{\bf 45}(3), pp.~636--639, 2020.

\bibitem{Fan}
H.~Fan, X.-W. Cao, L.~Wang, Z.-Z. Li, Q.-D. Chen, S.~Juodkazis, and H.-B. Sun,
  ``Control of diameter and numerical aperture of microlens by a single
  ultra-short laser pulse,'' {\em Optics Lett.}~{\bf 44}(21), pp.~5149--5152,
  2019.

\bibitem{Cao}
X.-W. Cao, Q.-D. Chen, L.~Zhang, Z.-N. Tian, Q.-K. Li, L.~Wang, S.~Juodkazis,
  and H.-B. Sun, ``Single-pulse writing of a concave microlens array,'' {\em
  Optics Lett.}~{\bf 43}(4), pp.~831--834, 2018.

\bibitem{Cao1}
X.-W. Cao, Q.-D. Chen, H.~Fan, L.~Zhang, S.~Juodkazis, and H.-B. Sun,
  ``Liquid-assisted femtosecond laser precision-machining of silica,'' {\em
  Nanomaterials}~{\bf 8}(5), p.~287, 2018.

\bibitem{16lsa16133}
M.~Malinauskas, A.~\v{Z}ukauskas, S.~Hasegawa, Y.~Hayasaki, V.~Mizeikis,
  R.~Buividas, and S.~Juodkazis, ``Ultrafast laser processing of materials:
  from science to industry,'' {\em Light: Sci. Appl.}~{\bf 5}(8),
  pp.~e16133--e16133, 2016.

\bibitem{Pancharatnam}
R.~Nityananda, ``The interference of polarised light: The pancharatnam phase,''
  {\em Resonance} (4), pp.~309--322, 2013.

\bibitem{98jjap1527}
M.~Watanabe, H.-B. Sun, S.~Juodkazis, T.~Takahashi, S.~Matsuo, Y.~Suzuki,
  J.~Nishii, and H.~Misawa, ``Three-dimensional optical data storage in
  vitreous silica,'' {\em Jpn. J. Appl. Phys.}~{\bf 27}(12B), pp.~L1527--L1530,
  1998.

\bibitem{Fan1}
H.~Fan, M.~Ryu, R.~Honda, J.~Morikawa, Z.-Z. Li, L.~Wang, J.~Maksimovic,
  S.~Juodkazis, Q.-D. Chen, and H.-B. Sun, ``Laser-inscribed stress-induced
  birefringence of sapphire,'' {\em Nanomaterials}~{\bf 9}(10), p.~1414, 2019.

\bibitem{new}
I.~Inoue, T.~Osaka, T.~Hara, T.~Tanaka, T.~Inagaki, T.~Fukui, S.~Goto,
  Y.~Inubushi, H.~Kimura, R.~Kinjo, H.~Ohashi, K.~Togawa, K.~Tono, M.~Yamaga,
  H.~Tanaka, T.~Ishikawa, and M.~Yabashi, ``Generation of narrow-band {X-ray}
  free-electron laser via reflection self-seeding,'' {\em Nat. Photonics}~{\bf
  13}(5), p.~319–322, 2019.

\bibitem{AV14}
A.~Vijayakumar and S.~Bhattacharya, ``Quasi-achromatic fresnel zone lens with
  ring focus,'' {\em Appl. Opt.}~{\bf 53}, pp.~1970--1974, 2014.

\bibitem{AV15}
A.~Vijayakumar and S.~Bhattacharya, ``Design of multifunctional diffractive
  optical elements,'' {\em Opt. Eng.}~{\bf 54}, p.~024104, 2014.

\bibitem{Descour99}
M.~R. Descour, D.~I. Simon, and W.-H. Yeh, ``Ring-toric lens for focus-error
  sensing in optical data storage,'' {\em Appl. Opt.}~{\bf 38}, pp.~1388--1392,
  1999.

\bibitem{GS}
R.~W. Gerchberg and W.~O. Saxton, ``A practical algorithm for the determination
  of phase from image and diffraction plane pictures,'' {\em Optik}~{\bf 35},
  p.~227–246, 1972.

\bibitem{Haist}
T.~Haist, M.~Sch\"{o}nleber, and H.~J. Tiziani, ``Computer-generated holograms
  from {3D}-objects written on twisted-nematic liquid crystal displays,'' {\em
  Opt. Commun.}~{\bf 140}, p.~299–308, 1997.

\bibitem{Piestun}
R.~Piestun, B.~Spektor, and J.~Shamir, ``Wave fields in three dimensions:
  analysis and synthesis,'' {\em J. Opt. Soc. Am. A}~{\bf 13}, pp.~1837--1848,
  1996.

\bibitem{Levy}
U.~Levy, D.~Mendlovic, Z.~Zalevsky, G.~Shabtay, and E.~Marom, ``Iterative
  algorithm for determining optimal beam profiles in a three-dimensional
  space,'' {\em Appl. Opt.}~{\bf 38}, pp.~6732--6736, 1999.

\bibitem{AV18}
M.~R. Rai, A.~Vijayakumar, and J.~Rosen, ``Extending the field of view by a
  scattering window in an {I-COACH} system,'' {\em Opt. Lett.}~{\bf 45}(5),
  pp.~1043--1046, 2018.

\bibitem{Arlt}
J.~Arlt and K.~Dholakia, ``Generation of high-order {Bessel} beams by use of an
  axicon,'' {\em Opt. Commun.}~{\bf 177}, p.~297–301, 2000.

\bibitem{Sun11}
Q.~Sun, K.~Zhou, G.~Fang, Z.~Liu, and S.~Liu, ``Generation of spiraling
  high-order {Bessel} beams,'' {\em Appl. Phys. B}~{\bf 104}, p.~215–221,
  2011.

\bibitem{16oe16988}
A.~Bal\v{c}ytis, D.~Hakobyan, M.~Gabalis, A.~\v{Z}ukauskas, D.~Urbonas,
  M.~Malinauskas, R.~Petru\v{s}kevi\v{c}ius, E.~Brasselet, and S.~Juodkazis,
  ``Hybrid curved nano-structured micro-optical elements,'' {\em Opt.
  Express}~{\bf 24}(15), pp.~16988 -- 16998, 2016.

\bibitem{AV152}
A.~Vijayakumar and S.~Bhattacharya, ``Compact generation of superposed
  higher-order {Bessel} beams via composite diffractive optical elements,''
  {\em Opt. Eng.}~{\bf 54}, p.~111310, 2015.

\bibitem{Bekshaev}
A.~Y. Bekshaev and M.~S. Soskin, ``Transverse energy flows in vectorial fields
  of paraxial beams with singularities,'' {\em Opt. Commun.}~{\bf 271}(2),
  p.~332–348, 2007.

\bibitem{Viswanathan}
V.~Kumar and N.~K. Viswanathan, ``Topological structures in the {Poynting}
  vector field: an experimental realization,'' {\em Opt. Lett.}~{\bf 38},
  pp.~3886--3889, 2013.

\bibitem{AV17}
A.~Vijayakumar, B.~Vinoth, I.~V. Minin, J.~Rosen, O.~V. Minin, and C.-J. Cheng,
  ``Experimental demonstration of square {Fresnel} zone plate with chiral side
  lobes,'' {\em Appl. Opt.}~{\bf 56}, pp.~F128--F133, 2017.

\bibitem{19n1263}
R.~Dharmavarapu, K.-I. Izumi, I.~Katayama, S.~Ng, J.~Vongsvivut, M.~J. Tobin,
  Y.~Nishijima, S.~Bhattacharya, and S.~Juodkazis, ``Dielectric cross-shaped
  resonator based metasurface for vortex beam generation in mid-ir and {THz}
  wavelengths,'' {\em Nanophotonics}~{\bf 8}(7), p.~1263–1270, 2019.

\bibitem{Pachava}
S.~Pachava, R.~Dharmavarapu, A.~Vijayakumar, S.~Jayakumar, A.~Manthalkar,
  A.~Dixit, N.~K. Viswanathan, B.~Srinivasan, and S.~Bhattacharya, ``Generation
  and decomposition of scalar and vector modes carrying orbital angular
  momentum: a review,'' {\em Opt. Eng.}~{\bf 59}(4), p.~041205, 2019.

\bibitem{Litvin}
I.~A. Litvin, A.~Dudley, and A.~Forbes, ``Poynting vector and orbital angular
  momentum density of superpositions of {Bessel} beams,'' {\em Opt.
  Express}~{\bf 19}, pp.~16760--16771, 2011.

\bibitem{Litvin14}
I.~A. Litvin, S.~Ngcobo, D.~Naidoo, K.~Ait-Ameur, and A.~Forbes, ``Doughnut
  laser beam as an incoherent superposition of two petal beams,'' {\em Opt.
  Lett.}~{\bf 39}, pp.~704--707, 2014.

\bibitem{AV19}
A.~Vijayakumar, C.~Rosales-Guzm\'{a}n, M.~R. Rai, J.~Rosen, O.~V. Minin, I.~V.
  Minin, and A.~Forbes, ``Generation of structured light by multilevel orbital
  angular momentum holograms,'' {\em Opt. Express}~{\bf 27}, pp.~6459--6470,
  2019.

\end{thebibliography}
\appendix
\renewcommand\thefigure{\thesection.\arabic{figure}}    
\setcounter{figure}{0}
\section*{Appendixes}

The optical configuration shown in Fig. 7 is analyzed by substituting different possible optical elements for OE1 and OE2 to study explosion and implosion in light-matter interactions.

\subsection*{Appendix 1 - Simulation of Optical Configuration of Fig. 7 with an axicon as OE1 and Fresnel axicon as OE2}

\begin{figure}[b]
\centering \includegraphics[width=16cm]{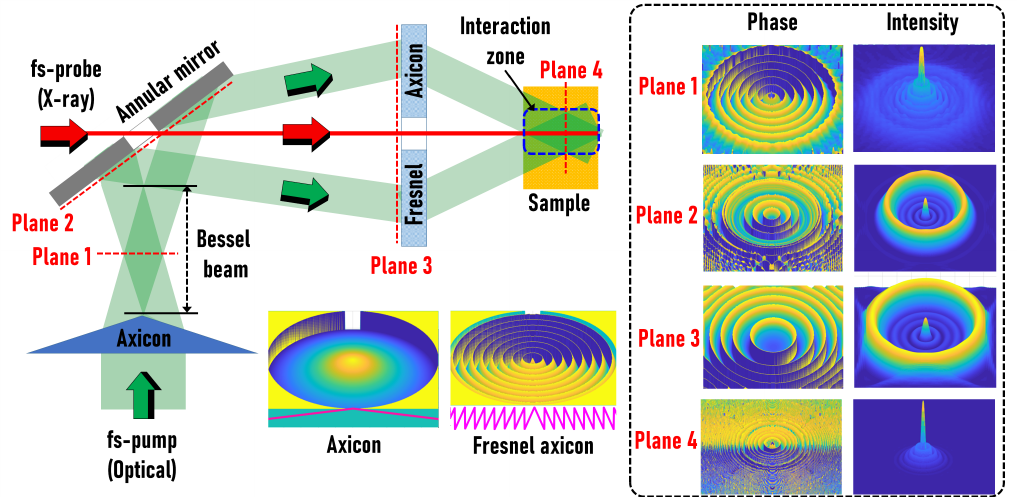}
\caption{Optical configuration of Fig. 7 with axicon as OE1 and Fresnel axicon as OE2. The simulation of the complex amplitude at four different planes of interest are shown.} \label{Figure_Appendix1}
\end{figure}

The optical configuration for studying light-matter interactions with co-propagating fs-pump (optical) and fs-probe (X-ray) is studied using a solid axicon as OE1 and a Fresnel axicon with a central hole (Fig. 3) as OE2 is shown in Fig.~\ref{Figure_Appendix1}. Collimated light with uniform illumination is incident on a solid axicon with base angle $\alpha$ and refractive index $n_a$. The phase of the solid axicon is given as $\Phi_A=(2\pi/\lambda)(n_a-1)\alpha R$, where $R=\sqrt{x^2+y^2}$ is the radial coordinate. The complex amplitude at a distance $z$ can be expressed as a convolution of the complex amplitude from the axicon $exp(-j\Phi_A$)  with the quadratic phase function $exp(j \pi R^2/\lambda z)$ corresponding to that distance $z$. The intensity at any plane $z$ is $I=|exp(-j\Phi_A) \otimes exp(j \pi R^2/\lambda z)|^2$, where '$\otimes$' is a 2D convolutional operator. Four planes are considered for the calculation of the complex amplitude. The first plane is selected to be within the focal depth of the axicon. The simulated phase profile shows a conical phase and the intensity has a Bessel-like profile as shown in Fig.~\ref{Figure_Appendix1}. The planes 2 and 3 are selected beyond the focal depth of the axicon with the plane 2 matching with the annular mirror plane and the plane 3 matching with the location of the Fresnel axicon\cite{VijayakumarSPIE,Golub}. The variation of intensity profile between the two planes shows the evolution of the ring pattern which is the characteristic of an axicon\cite{Vijayakumar} as the conical phase expands. The angle introduced by the annular mirror is avoided in the simulation. The ring pattern encounters the Fresnel axicon whose complex amplitude is given as $exp(-j2 \pi R/\Lambda)$. Therefore, the complex amplitude immediately after plane 3 is given as $E=exp(-j\Phi_A) \otimes exp(j \pi R^2/\lambda z) \times exp(-j2 \pi R/\Lambda)$.  The expression can be rewritten as $E=exp\{-j(\Phi_A+2 \pi R/\Lambda)\} \otimes exp(j \pi R^2/\lambda z)$. The two axicon phases generate a conical phase with a larger base angle and the complex amplitude at a plane closer to the interaction plane resembles a compressed Bessel-like beam with a shorter focal depth and sharper peak.

\begin{figure}[tb]
\centering \includegraphics[width=16cm]{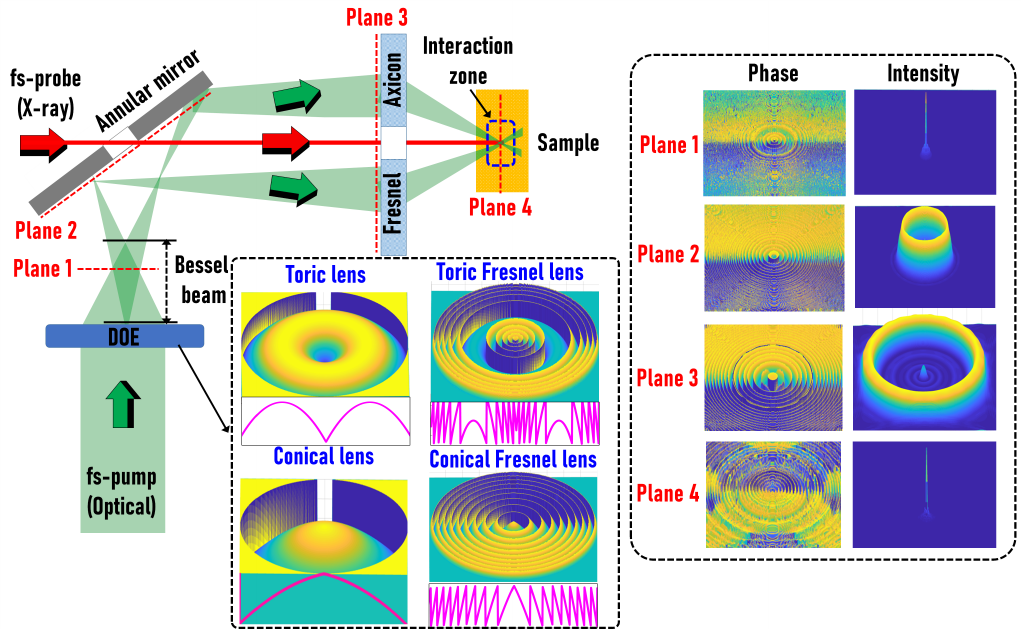}
\caption{Optical configuration of Fig. 7 with conical or toric lens as OE1 and Fresnel axicon as OE2. The simulation of the complex amplitude at four different planes of interest are shown.} \label{Figure_Appendix2}
\end{figure}

The advantages and disadvantages of employing a solid axicon as OE1 are as follows. The solid axicon is an optical element with a simpler shape function which is easier for fabrication and less expensive. The complex amplitude generated by the solid axicon-Fresnel axicon pair near the interaction zone is ideal. The use of solid axicon concentrates light in a ring in plane 3 where the Fresnel axicon is located and so the fabrication area of the Fresnel axicon can be much smaller, reducing the cost and fabrication time. However, there are few drawbacks. The intensity distribution has a peak at the center which will not be modulated by the Fresnel axicon as it overlaps with the central hole and will not contribute to the Bessel beam formation. Secondly, the thickness of the ring pattern is larger at plane 2 which demands a larger annular mirror.   

\begin{figure}[tb]
\centering \includegraphics[width=16cm]{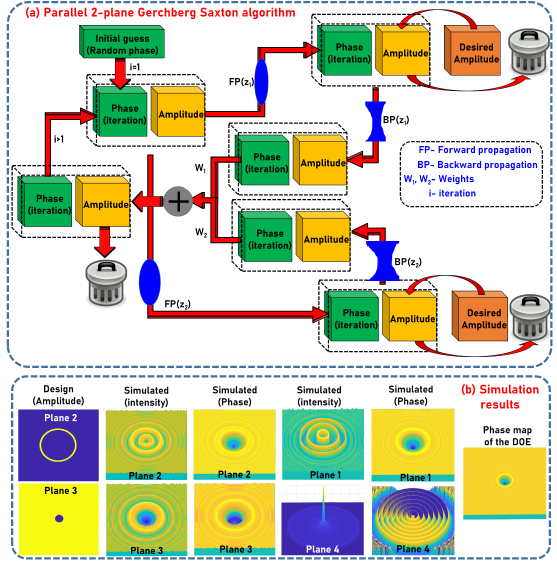}
\caption{(a) Flow chart of the two plane GSA and (b) the simulation results of the complex amplitude at four different planes of interest. } \label{Figure_Appendix3}
\end{figure}

\subsection*{Appendix 2 - Simulation of Optical Configuration of Fig. 7 with a conical or toric lens as OE1 and Fresnel axicon as OE2}

Conical and toric lenses can focus a uniform illumination to a ring and are used for various applications~\cite{AV14,AV15,Descour99}. Since the ring pattern obtained from a toric lens or a conical lens is focused unlike an axicon, a smaller annular mirror could be used. The use of a conical or toric lens also suppresses the central diffraction order. The optical configuration of Fig. 7 with a conical or toric lens is shown in Fig.~\ref{Figure_Appendix2}. The presence of an additional quadratic phase in this case compresses the area of the spread of Bessel beam resulting in a reduced interaction area near the sample but with a higher concentration of light energy. Both toric and conical lens has a relatively smoother shape function and therefore can be easily fabricated.     
\begin{figure}[tb]
\centering \includegraphics[width=16cm]{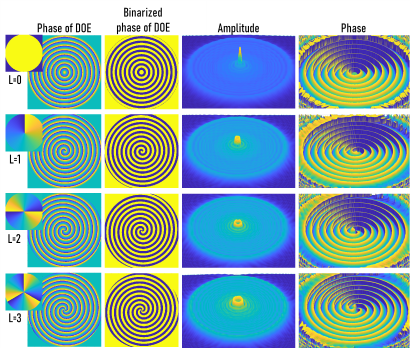}
\caption{Phase images of the spiral axicon as modulo-$2 \pi$ and binary [0,$\pi$] for $L$=0,1,2 and 3, and the corresponding complex amplitude images.} \label{Figure_Appendix4}
\end{figure}

\subsection*{Appendix 3 - Simulation of Optical Configuration of Fig. 7 with a special optical element as OE1 and Fresnel axicon as OE2}

In this case, an element is designed using the Gerchberg-Saxton algorithm (GSA). In the optical configuration, there are two planes of interest, namely plane 2 and plane 3. The original GSA was developed for synthesizing a pure phase function to generate a desired intensity distribution in a particular axial plane, usually the Fourier plane~\cite{GS}. A modified GSA was developed using Fresnel propagators to synthesize phase patterns that can generate desired intensity distributions in more than one plane in parallel~\cite{Haist,Piestun,Levy}. In this study, the GSA multiple plane approach was implemented to generate desired intensity profiles at plane 2 and plane 3. The flow chart of the two plane GSA is shown in Fig.~\ref{Figure_Appendix3}(a). The phase retrieval process for two planes is summarized here. The algorithm begins with an initial random guess of the phase profile. The initial complex amplitude is forward propagated using scalar diffraction formula to two different axial planes and the magnitudes of the complex amplitudes are replaced by the desired ones. The modified complex amplitudes are back propagated to the respective planes and summed. The phase of the resulting complex amplitude is extracted for the next iteration and so on until a minimum error value is obtained between the desired and the obtained intensity distributions. There are weighting factors $W1$ and $W2$ attached to the two complex amplitudes before they are added. In this way, it is possible to select a plane where the intensity distribution is more critical than the other and give additional weight.  The two intensity profiles designed for this study are a sharp ring pattern at plane 2 and a constant intensity with a central zero intensity at plane 3~\cite{AV18}. The intensity and phase of the complex amplitudes at different planes are shown in Fig.~\ref{Figure_Appendix3}(b). Unlike the previous cases, the central diffraction peak is close to zero in this case. In addition, the phase profile is symmetric and the phase variation is less than $\pi$ which makes the fabrication easier. The complex amplitude near the interaction zone has a larger axial interaction length and Bessel-like intensity profile. In comparison to the above two methods, the results obtained from the two plane GSA is better and does not suffer from fabrication challenges arising from random profiles as in usual cases of GSA.   

\begin{figure}[tb]
\centering\includegraphics[width=16cm]{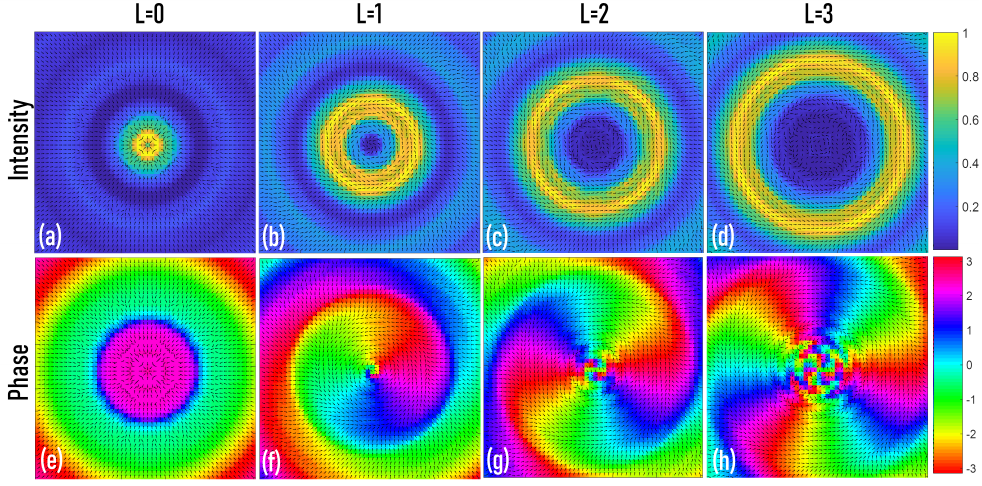}
\caption{The Poynting vector plots superposed on the normalised intensity images for (a) $L$=0, (b) $L$=1, (c) $L$=2, (d) $L$=3 and phase images [$-\pi,\pi$] (e) $L$=0, (f) $L$=1, (g) $L$=2 and (h) $L$=3 simulated at plane 4 close to the interaction zone. } \label{Figure_Appendix5}
\end{figure}

\subsection*{Appendix 4 - Simulation of Optical Configuration of Fig. 7 with a special optical element as OE1 and spiral axicon as OE2}

In this case, the special optical element synthesized from the two plane GSA is retained while the Fresnel axicon is replaced by a spiral axicon for OE2. The complex amplitude of the spiral axicon is given as $exp(-j2 \pi R/\Lambda) \times exp(j L\theta)$, where $L$ is the topological charge and $\theta$ is the azimuthal angle. The spiral axicon is capable of generating higher order Bessel beams (HOBB) with a large focal depth and carries orbital angular momentum~\cite{Arlt,Sun11,16oe16988,AV152}. The images of the modulo-2$\pi$ phase of the spiral axicon for $L$=0, 1, 2 and 3 are shown in Fig.~\ref{Figure_Appendix4}. The binary version [0,$\pi$] is shown which can be fabricated similar to the results shown in Fig. 3. The complex amplitude at plane 4 can be estimated as $exp(-j2 \pi R/\Lambda) \times exp(j L\theta) \otimes exp(j \pi R^2/\lambda z)$. The expression can be rearranged as $ exp(j L\theta)\times \{exp(-j2 \pi R/\Lambda) \otimes exp(j \pi R^2/\lambda z)\}$. It is well-established that a convolution between the complex amplitudes of an axicon and a quadratic phase function results in a Bessel-like complex amplitude. This is described in the main text as well. Therefore, the rearrangement of the complex amplitudes with convolution and product shows that the Bessel-like complex amplitude is multiplied by the complex amplitude of the spiral phase function. Consequently, the introduction of the spiral phase creates phase singularities in the Bessel like complex amplitudes resulting in the formation of HOBB. It is seen in Fig.~\ref{Figure_Appendix4} that the radius of the doughnut increases with the topological charge. The amplitude and phase for $L$=0, 1, 2 and 3 are shown in Fig.~\ref{Figure_Appendix4}.

Poynting vector plot is a direct method to understanding the directional energy flux or the flow of field. Since all the fields that are studied here are scalar fields, the Poynting vector can be calculated as the product of the intensity and the gradient of the phase~\cite{Bekshaev,Viswanathan}. The Poynting vector is given as $S_{x,y}\propto I_{x,y} \Delta\Phi_{x,y}$. The Poynting vector plot is calculated for four cases namely $L$ = 0 (Fresnel axicon), 1, 2 and 3 (Spiral axicon) at a plane within the interaction zone. The Poynting vector plots superposed on the respective intensity and phase distributions for the four cases are shown in Fig.~\ref{Figure_Appendix5}. From the figure, it is seen that in the case of Fresnel axicon ($L$=0), the Bessel beam can still exert radiation pressure forces due to the intensity gradients~\cite{AV17}. For other cases, as $L\neq 0$, it is seen that the flux spirals around the center indicating the presence of the orbital angular momentum which can exert forces. When $L$ increases, the force field also increases. It is also possible to generate hybrid force fields with both forces from radiation pressure and also orbital angular momentum by spatial multiplexing and by manipulating the phase values. 

\subsection*{Summary on possible realisation of optical fs-pump}

In this perspective study, we attempted to realize the optical configuration of Fig. 7 using basic optical phase functions such as conical phase, spiral phase, quadratic phase, toric phase and retrieved phase using GSA. These fundamental theoretical and numerical studies reveal the feasibility of the explosion and implosion studies in light-matter interactions with basic optical elements and optical engineering. However, experimental studies are needed to understand the impact of the force fields in the proposed optical configurations. The recent developments on the generation of exotic optical fields such as fields varying in amplitude, phase and polarization~\cite{19n1263,Pachava}, vector fields~\cite{Pachava}, fields generated by OAM superposition~\cite{Litvin,Litvin14,AV152}, OAM beams in unconventional beam channels~\cite{AV19} will enable sophisticated light-matter interactions studies.        

\end{document}